\documentclass[12pt]{article}
\usepackage[utf8]{inputenc}

\usepackage{tikz}
\definecolor{colorSFROLAIC}{HTML}{cc0000}
\definecolor{colorSFROLBIC}{HTML}{f1c232}
\definecolor{colorROL}{HTML}{6aa84f}
\definecolor{colorPROL}{HTML}{3d85c6}

\usepackage[unicode,linktoc=all,hidelinks]{hyperref}
\usepackage[a4paper, margin=2cm]{geometry}
\usepackage{mathtools}
\usepackage{amsmath}
\usepackage{amssymb}
\usepackage{amsthm}
\usepackage{bm}
\usepackage{bbm}

\usepackage{float} 
\usepackage{adjustbox}
\usepackage{threeparttable}
\usepackage{booktabs}

\usepackage{algorithm}
\usepackage{algorithmic}

\usepackage[english]{babel}
\usepackage{natbib}

\title{Joint Learning from Heterogeneous Rank Data}
\author{Sjoerd Hermes$^{1,2}$, Joost van Heerwaarden$^{1,2}$ and Pariya Behrouzi$^1$}
\date{%
    $^1$ Mathematical and Statistical Methods, Wageningen University\\%
    $^2$ Plant Production Systems, Wageningen University\\
}

\begin{document}

\maketitle
\begin{abstract}
\noindent The rank-ordered logit (ROL) model is frequently used to aggregate rankings from multiple individuals into a single ranking that corresponds to the unobserved utility derived from the ranked alternatives. When these individuals exhibit heterogeneous preferences, they can be grouped to reduce bias. However, the common practice of fitting ROL models separately for each group may overlook shared patterns across groups, resulting in a loss of information. We introduce the sparse fused rank-ordered logit (SFROL) model, an extension of the ROL model, that allows joint learning from heterogeneous ranking data, whereby information from different groups is utilised to achieve better model performance. In this framework, the observed rankings are modeled as a function of covariates pertaining to the ranked alternatives. By imposing penalties on the likelihood function, we allow for both the sharing of information on the corresponding coefficients across groups and the shrinkage of the coefficients to zero to improve interpretability, point estimation and prediction of the model. Simulations studies generally indicate superior performance of the proposed method compared to exsiting approaches across a wide variety of scenarios. Usage and interpretation of the method is illustrated on a sweet potato consumer preference application. An R package containing the proposed methodology can be found on \url{https://CRAN.R-project.org/package=SFPL.}
\end{abstract}

\noindent%
{\it Keywords:} Ranked data, rank-ordered logit, penalized likelihood, heterogeneity, grouped choice data.

\section{Introduction}
\noindent The statistical modelling of ranking data has important applications in a  variety of fields, such as economics (Beggs et al., \citeyear{beggs1981assessing}), education (Hassan \& Geys, \citeyear{hassan2017we}), marketing (Müllensiefen et al., \citeyear{mullensiefen2018using}) and transport (Hossain et al., \citeyear{hossain2022modeling}). Nevertheless, its origin lies in psychology, with the introduction of the law of comparative judgements by Thurstone (\citeyear{thurstone1927law}). This seminal paper has led to the rise of a wide variety of statistical models that aim to aggregate rankings made by individuals into a single ranking that corresponds to the underlying utility of the ranked alternatives. Aside from the Thurstonian model, some well-known models are the Babington Smith (Babington Smith, \citeyear{babingtonsmith1950}), Bradley-Terry (Bradley \& Terry, \citeyear{bradley1952rank}), Mallows (Mallows, \citeyear{mallows1957non}) and Plackett-Luce (Luce, \citeyear{luce1959}; Plackett, \citeyear{plackett1975analysis}) models, where the latter model is also known as the rank-ordered logit model (ROL) (Beggs et al., \citeyear{beggs1981assessing}; Allison \& Christakis, \citeyear{allison1994logit}). In this article, the ROL model is further explored, as it lends itself naturally to the statistical modelling of partial rankings, which is the type of data that we focus on. 

The ROL can be viewed as a generalization of the (multinomial) logit model for discrete choice sets (Ben-Akiva et al., \citeyear{ben1997modeling}), where this latter model requires data in which various individuals express their preference for one alternative over others from a discrete choice set. Conversely, the ROL can handle data in which individuals express their preferences by means of a rank ordering of the various alternatives comprising the discrete choice set. Typically, the observed rankings are assumed to arise from unobserved utilities that individuals derive from the alternatives, where a higher utility for a given alternative corresponds to a higher ranking. 

Even though the ROL model is broadly applicable to rank-ordered preference data on discrete choice sets, given its ability to handle partial rankings, its performance can be suboptimal in some scenarios. The scenario we consider here is one where individuals have heterogeneous preferences. Typically, real world data is not homogeneous: not all observations within the data are expected to originate from the same distribution. However, unlike existing approaches that aim to model heterogeneous preferences separately using the ROL model such as the methods proposed by Fok et al.\ (\citeyear{fok2012rank}) and Cheng et al.\ (\citeyear{cheng2024bayesian}), the method proposed here assumes that multiple known groups of individuals exist within the data, where more or less similar preference patterns can be found within groups, but where the preference patterns differ between groups. Differing preference patterns frequently occur in, for example, marketing, where the practice of market segmentation aims to divide a potential consumer base into subgroups of consumers, based on some features (or preferences) of these consumers (Denizci Guillet \& Kucukusta, \citeyear{denizci2016spa}). When individuals can be meaningfully grouped (based on prior knowledge, experimental design, or substantive interest) it is common practice to estimate a separate model for each group. This approach helps mitigate the bias that can arise when pooling heterogeneous preferences into a single model. Importantly, it assumes that preferences within each group are relatively homogeneous. However, estimating separate ROL models for each group ignores potential similarities across groups, which may lead to inefficient use of data, particularly when some covariate effects are shared.
 To the best of our knowledge, no work is done on efficient estimation of ROL models (or other ranking models) on data consisting of multiple known groups of individuals, where within a group the individuals have more or less similar preference patterns, but where these preference patterns differ between groups. As such, we propose a novel extension of the ROL model: the sparse fused rank-ordered logit model (SFROL), that allows for joint learning from multi-group data. By borrowing information across known groups, more information is used, which in turn can lead to improved parameter estimates (Danaher et al., \citeyear{danaher2014joint}; Dondelinger \& Mukherjee, \citeyear{dondelinger2020joint}; Hermes et al., \citeyear{hermes2024copula}). Information sharing across known groups has been introduced before for the Gaussian graphical model (Danaher et al., \citeyear{danaher2014joint}) and the multivariate regression model (Dondelinger \& Mukherjee \citeyear{dondelinger2020joint}). In line with these authors, we will penalise the likelihood to enforce joint learning of the model parameters. 
In addition to efficient joint learning from multi-group rank data, the proposed method makes use of covariates. These covariates describe some property of the ranked alternatives that may affect their utility. Consequently, the utilities are modelled as a function of these covariates, whereby the effects of each covariate can be group-specific. When the number of covariates is large, some form of variable selection is required, both because preference is likely determined by relatively few variables (Naik et al., \citeyear{naik2008challenges}; Zhu \& Liu \citeyear{zhu2023learning}) and because reliable parameter estimates becomes problematic as the number of variables approaches or exceeds the number of observations. For this reason, additional penalisation is introduced that forces coefficients to zero, depending on the chosen value of a tuning parameter, such that bias is induced to reduce the variance. Even when the data are not high-dimensional, this shrinkage provides a predictive advantage by avoiding model overfitting. This is especially valuable in situations where rankings are difficult or costly to obtain, but information for the alternatives is ubiquitous; researchers can fit the proposed method on the items for which ranking data exist, and predict the rankings of the other alternatives using the available covariates, whilst reducing the bias that overfitting would accrue. An example for such costly to obtain data can be found in the context of medical science, where the grouping of individuals providing rankings occurs due to disease states (see e.g.\ Mollica \& Tardella, \citeyear{mollica2014epitope}). Another example can be found in the field of marketing, which perhaps does not suffer from a low number of observations, but where instead both data and covariates are more ubiquitous. Here, the proposed method can be used to predict consumer response to various advertising campaigns, whenever different target groups are approached (Hariharan et al., \citeyear{hariharan2015optimal}; Shin \& Yu, \citeyear{shin2021targeted}). Even though shrinkage of coefficients has not been applied in ROL models before, it has been imposed on Bradley-Terry models (Schaumberger \& Tutz, \citeyear{schauberger2017subject}; \citeyear{schauberger2019btllasso}). Moreover, Jeon  and Choi (\citeyear{jeon2018sparse}) imposed a truncated lasso penalty on the ROL model, that fuses the utilities of similar alternatives together. Nevertheless, the fused-type group penalty has not been applied in ranking models before, nor have existing methods generalised the fitted model to predictive settings or demonstrated the model performance in terms of recovering the underlying rankings or coefficients.
\\~\\
\noindent The paper starts with a description of the ROL model, before extending it to obtain the proposed method in Section \ref{Methodology}. Parameter estimation of the model is discussed in Section \ref{Parameter estimation}. Section \ref{Simulation study} consists of a simulation study that evaluates the performance of the proposed method in terms of rank aggregation accuracy and coefficient estimation accuracy. An application of the proposed method on preference data pertaining to sweet potatoes is provided in Section \ref{Real World Data}. Finally, Section \ref{Conclusion} concludes the article and discusses some points of (future) interest.

\section{Methodology} \label{Methodology}
\noindent Before introducing the ROL model, some notation is required. Assume that we are given $K$ groups of ranking data $\bm{\Pi}^{(1)},\ldots,\bm{\Pi}^{(K)}$, where $\bm{\Pi}^{(k)} \in \mathbb{N}^{n_{k} \times m}$, $1 \leq k \leq K$, consists of $n_k$ individuals with preference patterns that are i.i.d., in which individual $i$ ranks $m$ alternatives from a total choice set of $M$ alternatives, with $m \leq M$. Conversely, individuals contained in different groups $k$ and $k'$ may exhibit preference patterns that do not arise from the same distribution. The $\bm{\pi}^{(k)}_i = \left(\pi^{(k)}_{i1},\ldots, \pi^{(k)}_{im}\right)$, $1 \leq i \leq n_k,$ are permutations of $1,\ldots,m$, such that $\pi^{(k)}_{ij}$ denotes the rank of alternative $j$ according to individual $i$ in group $k$. Any observed ranking given by individual $i$, $\bm{\pi}^{(k)}_i$, corresponds to an ordering of the unobserved utilities $\bm{u}^{(k)}_i = \left(u^{(k)}_{i1},\ldots, u^{(k)}_{im}\right)$ that individual $i$ has for the $m$ alternatives. Consequently, we have that $\bm{\pi}^{{(k)}^{-1}}_i = \bm{\sigma}^{(k)}_i = \left(\sigma^{(k)}_{i1},\ldots, \sigma^{(k)}_{im}\right)$, where $\sigma^{(k)}_{ij}$ indicates the alternative ranked $j$-th by individual $i$ in group $k$. Let the set of alternatives ranked by individual $i$ in group $k$ be denoted as as $\mathcal{O}_i^{(k)}$, where $\mathcal{O}_i^{(k)}$ need not equal $\mathcal{O}_{i'}^{(k)}$ for $i \neq i'$, although we do have that $\bigcup_{i=1}^{n_k}\mathcal{O}_i^{(k)} = \bm{\mathcal{O}}$ for all $k$, with $|\bm{\mathcal{O}}| = M$. In addition, we might have access to data consisting of $p$ covariates: $\bm{X} \in \mathbb{R}^{M \times p}$. These covariates provide information about some properties of the alternatives under consideration. Even though we assume that each individual ranks the same number of alternatives for notational simplicity, the proposed method holds for the more general setting consisting of a different number of ranked alternatives per individual.

The ROL postulates that the utility of alternative $j$ for individual $i$ in group $k$ is given by
\begin{equation}
\label{eq:utlilty}
	u_{ij}^{(k)} = \bm{x}_j\bm{\beta}^{(k)} + v^{(k)}_{ij},
\end{equation}
where the $v^{(k)}_{ij}$ are Gumbel distributed errors. The superscript $(k)$ implies that the the coefficients, and therefore also the utilities may vary per group, as a consequence of the differing preference patterns between groups. 

Given Equation \ref{eq:utlilty}, the probability of observing ranking $\bm{\pi}^{(k)}_i$ is given by
\begin{equation}
\label{eq:probrank}
    \mathbb{P}\left(\bm{\pi}^{(k)}_i |\bm{\beta}^{(k)}\right) = \prod_{j = 1}^{m}\frac{\exp\left(\bm{x}_{\sigma^{(k)}_{ij}}\bm{\beta}^{(k)}\right)}{\sum_{l = j}^m \exp\left(\bm{x}_{\sigma^{(k)}_{il}}\bm{\beta}^{(k)}\right)},
\end{equation}
where $\bm{x}_{\sigma^{(k)}_{ij}}$ denotes the vector of covariates corresponding to the alternative ranked $j$-th by individual $i$ in group $k$. Whenever the number of included covariates is large, some of them might be expected to have no contribution to an individual's relative preference for an alternative. Forcing small but nonzero coefficients to zero could therefore greatly improve interpretation, in addition to increasing the (predictive) robustness of the model. This forcing of some coefficients to zero is called shrinkage and is often done by penalising the maximum likelihood estimates of the parameter of interest (Hastie et al., \citeyear{hastie2015statistical}). A second penalty term can be added to the likelihood when the goal is to enforce information sharing between different known subgroups. Whenever different groups in the data have some commonalities, either in their preference patterns or in the coefficients governing the preference patterns, information sharing improves parameter estimates (Danaher et al., \citeyear{danaher2014joint}; Dondelinger \& Mukherjee, \citeyear{dondelinger2020joint}; Hermes et al., \citeyear{hermes2024copula}). We propose these shrinkage and information sharing processes by means of the following Equations
\begin{equation}
\label{eq:pen_beta}
\mathcal{P}(\bm{B}) = \lambda_s\sum_{k=1}^K\left|\left|\bm{\beta}^{(k)}\right|\right|_1 + \lambda_f\sum_{k<k'}\left|\left|\bm{\beta}^{(k)} - \bm{\beta}^{(k')}\right|\right|_1,
\end{equation}
where $\lambda_s \in \Lambda_s$ determines the level of shrinkage for the coefficients and $\lambda_f \in \Lambda_f$ determines the overall level of information sharing (fusion) between groups. These penalty functions are subtracted from the likelihood, such that larger values of $\lambda_s$ and $\lambda_f$ result in more shrinkage and more information sharing between groups respectively, whereas for $\lambda_s = \lambda_f = 0$ unpenalised maximum likelihood estimates are obtained. 

Assuming that obtained ranking data consists of $K$ known subgroups, where within a group $n_k$ i.i.d.\ full or partial rankings are observed, we obtain the following penalised negative log likelihood
\begin{equation}
\label{eq:penlik}
\begin{gathered}
\ell_\mathcal{P}(\bm{B}) = \sum_{k=1}^K\sum_{i = 1}^{n_k}\sum_{j = 1}^{m}\left\{\log\left[\sum_{l = j}^m \exp\left(\bm{x}_{\sigma^{(k)}_{il}}\bm{\beta}^{(k)}\right)\right] - \bm{x}_{\sigma^{(k)}_{ij}}\bm{\beta}^{(k)}\right\} \\
+  \lambda_s\sum_{k=1}^K\sum_{q=1}^p\left|\beta_q^{(k)}\right| + \lambda_f\sum_{k<k'}\sum_{q=1}^p\left|\beta_q^{(k)} - \beta_q^{(k')}\right|,
\end{gathered}
\end{equation}
which is convex in $\bm{B}$. A proof of convexity for a similar (unpenalised) problem can be found in Schäfer and Hüllermeister (\citeyear{schafer2018dyad}), whose extension to Equation (\ref{eq:penlik}) is straightforward, due to the convexity of the imposed penalty terms. Convexity is a desired property here, as many the performance of many iterative optimization methods is contingent upon convexity of the function. 
\\
\\
Before moving on to the parameter estimation, the concept of rank prediction ought to be discussed. Without covariates, the ROL is unable to predict the rank of unseen alternatives. However, the method proposed in this article does incorporate covariates. As such, contingent on that the same set of $p$ covariates for unseen alternative $M+1$ is provided, the estimated utility of that alternative for any group $k$ is given by $\hat{u}_{M+1}^{(k)} = \exp\left(\bm{x}_{M+1}\hat{\bm{\beta}}^{(k)}\right)$, where the rank of $M+1$ is given by the value of $\hat{u}_{M+1}^{(k)}$, relative to the earlier estimated $\hat{u}_{1}^{(k)},\ldots,\hat{u}_{M}^{(k)}$. These rankings depend on the group whose estimated coefficients are used, as for the new alternative too, different groups imply different preferences.  

Rank prediction directly relates to an important property of the ROL model: the independence of irrelevant alternatives (Luce, \citeyear{luce1959}). Removing more alternatives from the data on which the model is fitted does not change the relative rank ordering for the remaining alternatives in this data, due to irrelevance of alternatives. That is, if someone prefers alternative $j$ over $j'$, then if $j''$ is introduced, it cannot be that $j'$ is preferred over $j$ by this same person. Therefore, the unseen alternatives do not change the preference order of the seen alternatives, but when multiple unseen alternatives are predicted, their relative ordering might change compared to the true order, as we have no ranking data on them, only covariates. Consequently, the validity of using a ROL model with covariates in a predictive setting is determined by the correctness of the assumption on whether or not the one-way deterministic relationship between the covariates and the rankings is applicable to unseen alternatives. 

\section{Parameter estimation} \label{Parameter estimation}
\noindent Maximum likelihood estimation permits no analytic solution for the coefficients. Therefore, research centring around ROL models resort to iterative methods, such as the Alternating Directions Method of Multipliers (Yıldız et al., \citeyear{yildiz2020fast}), Generalized Method-of-Moments (Azari Soufiani et al., \citeyear{azari2013generalized}), Newton-Rhapson (Beggs et al., \citeyear{beggs1981assessing}), Majorization-Minimization (Hunter, \citeyear{hunter2004mm}; Jeon \& Choi, \citeyear{jeon2018sparse}) and Stochastic Gradient Descent (Cheng et al., \citeyear{cheng2010label}) approaches for maximum likelihood estimation. Without the inclusion of penalty terms in Equation (\ref{eq:pen_beta}), we could use the Newton–Raphson method to conduct maximum likelihood estimation. However, these penalty terms do not have continuous second order derivatives, and hence cause typical gradient-based methods to be unsuitable for this problem. To overcome this problem, we utlise the majorize–minimize (MM) algorithm, where a surrogate function $S$ is introduced that is quadratic, and majorizes (bounds) the convex objective function from above, and hence easier to optimize than the function of interest $\mathcal{P}$, see Equation (\ref{eq:pen_beta}). The algorithm minimizes the surrogate function iteratively, and given an estimate for iteration $h$, $\hat{\bm{B}}^{[h]}$, the surrogate function has the property that $\mathcal{P}(\hat{\bm{B}}^{[h]}) = S(\hat{\bm{B}}^{[h]}|\hat{\bm{B}}^{[h]})$ and $\mathcal{P}(\bm{B}) \leq S(\bm{B}|\hat{\bm{B}}^{[h]})$. As the MM algorithm has the property that $\mathcal{P}\left(\hat{\bm{B}}^{[h+1]}\right) \leq \mathcal{P}\left(\hat{\bm{B}}^{[h]}\right)$, our aim is to find a $\bm{B}$ that minimizes $S(\bm{B}|\hat{\bm{B}}^{[h]})$. Slightly modifying the surrogate function proposed by Yu et al.\ (\citeyear{yu2015high}) for a similar problem gives us the following surrogate function for Equation (\ref{eq:pen_beta})
\begin{equation*}
\begin{gathered}
    S\left(\bm{B}|\hat{\bm{B}}^{[h]}\right) = \lambda_s\sum_{k=1}^K\sum_{q=1}^p\left[\left|\hat{\beta}_q^{(k)^{[h]}}\right| - \epsilon\log\left(1 + \frac{\left|\hat{\beta}_q^{(k)^{[h]}}\right|}{\epsilon}\right) + \frac{\left(\beta^{(k)}_q\right)^2 - \left(\hat{\beta}_q^{(k)^{[h]}}\right)^2}{2\left(\left|\hat{\beta}_q^{(k)^{[h]}}\right| + \epsilon\right)}\right]\\
        \begin{split}
     + \lambda_f\sum_{k<k'}\sum_{q=1}^p\left[\left|\hat{\beta}_q^{(k)^{[h]}} - \hat{\beta}_q^{(k')^{[h]}}\right| - \epsilon\log\left(1 + \frac{\left|\hat{\beta}_q^{(k)^{[h]}} - \hat{\beta}_q^{(k')^{[h]}}\right|}{\epsilon}\right)\right. \\
     \left. + \frac{\left(\beta_q^{(k)} - \beta_q^{(k')}\right)^2 - \left(\hat{\beta}_q^{(k)^{[h]}} - \hat{\beta}_q^{(k')^{[h]}}\right)^2}{2\left(\left|\hat{\beta}_q^{(k)^{[h]}} - \hat{\beta}_q^{(k')^{[h]}}\right| + \epsilon\right)}\right],
    \end{split}
\end{gathered}
\end{equation*} 
where we set $\epsilon = 10^{-5}$ (c.f.\ Hunter \& Li, \citeyear{hunter2005variable}; Yu et al., \citeyear{yu2015high}). We can then combine this surrogate function with the non-penalised part of the negative log likelihood of Equation (\ref{eq:penlik}) to obtain a surrogate penalised log likelihood 
\begin{equation}
\label{eq:surrogate_pen_loglik}
Q\left(\bm{B}|\hat{\bm{B}}^{[h]}\right) = \ell(\bm{B}) + S\left(\bm{B}|\hat{\bm{B}}^{[h]}\right),
\end{equation} 
where $\ell(\bm{B}) = \ell_\mathcal{P}(\bm{B}) - \mathcal{P}(\bm{B})$. Observe that the surrogate function given by Equation (\ref{eq:surrogate_pen_loglik}) now has the properties that $\ell_\mathcal{P}(\hat{\bm{B}}^{[h]}) = Q(\hat{\bm{B}}^{[h]}|\hat{\bm{B}}^{[h]})$ and $\ell_\mathcal{P}(\bm{B}) \leq Q(\bm{B}|\hat{\bm{B}}^{[h]})$. Using the Newton–Raphson method, see Hunter and Li (\citeyear{hunter2005variable}), the parameter estimate for iteration $h+1$ is given by
\begin{gather}
\hat{\bm{B}}^{[h+1]} = \hat{\bm{B}}^{[h]} - u_h\left[\nabla^2  Q\left(\hat{\bm{B}}^{[h]}\right)\right]^{-1}\nabla Q\left(\hat{\bm{B}}^{[h]}\right)\notag\\
= \hat{\bm{B}}^{[h]} - u_h\left[\nabla^2  \ell\left(\hat{\bm{B}}^{[h]}\right) + \lambda_s\bm{V}_s^{[h]} + \lambda_f\bm{V}_f^{[h]}\right]^{-1}\left[\nabla \ell\left(\hat{\bm{B}}^{[h]}\right) + \left(\lambda_s\bm{V}_s^{[h]} + \lambda_f\bm{V}_f^{[h]}\right)\hat{\bm{B}}^{[h]}\right]\label{eq:update_step_b},
\end{gather}
for $u_h > 0$ and where both $\bm{V}_s^{[h]}$ and $\bm{V}_f^{[h]} \in \mathbb{R}^{pK \times pK}$ are block matrices. For some matrix $\bm{V} \in \mathbb{R}^{pK \times pK}$, let $v_{qk,q'k'}$ denote element $q,q'$ of block $k,k'$, resulting in 
\begin{equation*}
v_{f; qk,q'k'}^{[h]} =
  \begin{cases}
    \sum_{k < k'}\frac{1}{\left|\hat{\beta}_q^{(k)^{[h]}} - \hat{\beta}_{q'}^{(k')^{[h]}}\right| + \epsilon}, & \text{if } q = q' \text{ and } k = k' \\
    \frac{-1}{\left|\hat{\beta}_q^{(k)^{[h]}} - \hat{\beta}_{q'}^{(k')^{[h]}}\right| + \epsilon}, & \text{if } q = q' \text{ and } k < k'\\
    0, & \text{if } q \neq q' \text{ or } k > k'
  \end{cases}
\end{equation*} 
and
\begin{equation*}
v_{s; qk,q'k'}^{[h]} = 
\begin{cases}
\frac{1}{\left|\hat{\beta}_q^{(k)^{[h]}}\right| + \epsilon}, & \text{if } q = q' \text{ and } k = k' \\
0, & \text{otherwise}.
\end{cases}
\end{equation*}
A Newton–Raphson step is used to obtain the initial parameter estimate $\hat{\bm{B}}^{[0]}$, which is the MLE based on $\ell(\bm{B})$. The gradient $\nabla \ell\left(\hat{\bm{B}}^{[h]}\right) = \left(\nabla \ell\left(\hat{\bm{\beta}}^{(1)^{[h]}}\right),\ldots,\nabla \ell\left(\hat{\bm{\beta}}^{(K)^{[h]}}\right)\right)^T$ has the following analytic solution
\begin{equation*}
\nabla \ell\left(\hat{\bm{\beta}}^{(k)^{[h]}}\right) = \sum_{i = 1}^{n_k}\sum_{j = 1}^{m}\left\{\frac{\sum_{l = j}^m \bm{x}_{\sigma^{(k)}_{il}}\exp\left(\bm{x}_{\sigma^{(k)}_{il}}\hat{\bm{\beta}}^{(k)^{[h]}}\right)}{\sum_{l = j}^m \exp\left(\bm{x}_{\sigma^{(k)}_{il}}\hat{\bm{\beta}}^{(k)^{[h]}}\right)} - \bm{x}_{\sigma^{(k)}_{ij}}\right\},
\end{equation*}
due to the separability of $\ell(\bm{B})$ for the different $k$. Similarly, we can represent the Hessian $\nabla^2  \ell\left(\hat{\bm{B}}^{[h]}\right)$ by the following block matrix
\begin{equation*}
\nabla^2  \ell\left(\hat{\bm{B}}^{[h]}\right) = \begin{pmatrix}
\nabla^2 \ell\left(\hat{\bm{\beta}}^{(1)^{[h]}}\right) & \bm{0} & \hdots & \bm{0}\\
\bm{0} & \nabla^2 \ell\left(\hat{\bm{\beta}}^{(2)^{[h]}}\right) & \hdots & \bm{0}\\
\vdots & \vdots & \ddots & \vdots\\
\bm{0} & \bm{0} & \hdots & \nabla^2 \ell\left(\hat{\bm{\beta}}^{(K)^{[h]}}\right)
\end{pmatrix},
\end{equation*}
where each nonzero block has the following entries
\begin{equation*}
\begin{gathered}
\nabla^2 \ell\left(\hat{\bm{\beta}}^{(k)^{[h]}}\right) = \sum_{i = 1}^{n_k}\sum_{j = 1}^{m}\left\{\frac{\left[\sum_{l = j}^m\bm{x}_{\sigma^{(k)}_{il}}\bm{x}_{\sigma^{(k)}_{il}}^T\exp\left(\bm{x}_{\sigma^{(k)}_{il}}\hat{\bm{\beta}}^{(k)^{[h]}}\right)\right]\left[\sum_{l = j}^m\exp\left(\bm{x}_{\sigma^{(k)}_{il}}\hat{\bm{\beta}}^{(k)^{[h]}}\right)\right]}{\left[\sum_{l = j}^m \exp\left(\bm{x}_{\sigma^{(k)}_{il}}\hat{\bm{\beta}}^{(k)^{[h]}}\right)\right]^2}\right\}\\ 
- \sum_{i = 1}^{n_k}\sum_{j = 1}^{m}\left\{\frac{\left[\sum_{l = j}^m\bm{x}_{\sigma^{(k)}_{il}}\exp\left(\bm{x}_{\sigma^{(k)}_{il}}\hat{\bm{\beta}}^{(k)^{[h]}}\right)\right]\left[\sum_{l = j}^m\bm{x}_{\sigma^{(k)}_{il}}^T\exp\left(\bm{x}_{\sigma^{(k)}_{il}}\hat{\bm{\beta}}^{(k)^{[h]}}\right)\right]}{\left[\sum_{l = j}^m \exp\left(\bm{x}_{\sigma^{(k)}_{il}}\hat{\bm{\beta}}^{(k)^{[h]}}\right)\right]^2}\right\}.
\end{gathered}
\end{equation*}
Both the gradient and the Hessian can be plugged into Equation (\ref{eq:update_step_b}), which is repeatedly estimated for $h = 0,1,2,\ldots$ until
\begin{equation*}
\left|\frac{\ell_\mathcal{P}(\hat{\bm{B}}^{[h+1]}) - \ell_\mathcal{P}(\hat{\bm{B}}^{[h]})}{\ell_\mathcal{P}(\hat{\bm{B}}^{[h]})}\right| \leq \xi,
\end{equation*} 
for some small $\xi > 0$. Convergence results for this method are provided by both Hunter and Li (\citeyear{hunter2005variable}) and Yu et al.\ (\citeyear{yu2015high}), and will not be repeated here, as their results hold for general convex and differentiable likelihoods. 
\\~\\
\noindent The proposed method requires the selection of two penalty parameters: $\lambda_s$ and $\lambda_f$. Throughout this paper, these are selected in a data-driven fashion by means of 5-fold cross-validation, where the proposed method is fitted across a variety of values for $\lambda_s \in \Lambda_s$ and $\lambda_f \in \Lambda_f$. Subsequently, the combination of $\lambda_s$ and $\lambda_f$ that maximizes a score function reflecting the number of correctly estimated rankings (the Rank Correctness Ratio, see Section \ref{Simulation study}), averaged across all 5 folds is selected as the “optimal" penalty parameter combination. The grid of $|\Lambda_s \times \Lambda_f|$ penalty parameters is chosen such that for $\min{\Lambda_s}$ and $\min{\Lambda_f}$ we obtain the maximum likelihood estimates, whilst for $\max{\Lambda_s}$ we obtain $\bm{\beta}^{(k)} = \bm{0}$ for all $1 \leq k \leq K$ and for $\max{\Lambda_f}$ we obtain $\bm{\beta}^{(k)} = \bm{\beta}^{(k')}$ for all $k \neq k'$.
\\~\\
\noindent On a different note, despite the attractiveness of the ROL model with covariates, the total choice set $\bm{\mathcal{O}}$ considered in the ranking process should grow with the number of covariates $p$ to ensure identifiability of the $\bm{\beta}^{(k)}$. More formally, the ROL model is identifiable if and only if rank$(\bm{X}) = p$. This holds for both the proposed method, as well as for existing ROL models with covariates. Even though this is an important constraint on the model, it is not mentioned in the existing literature on ROL models with covariates (cf.\ Cheng et al., \citeyear{cheng2010label}; Schäfer \& Hüllermeister, \citeyear{schafer2018dyad}; Yıldız et al., \citeyear{yildiz2020fast}), potentially misleading practitioners into believing the usefulness of their analyses, whilst in reality the estimated $\bm{\beta}^{(k)}$ are non-unique if rank$(\bm{X}) = M < p$. A simple proof for this condition is provided in Appendix B.

\section{Simulation study} \label{Simulation study}
\noindent To evaluate the performance of the proposed method, both in absolute terms and relative to existing methods, a simulation study is conducted. In this simulation study, we evaluate how the methods perform in terms of recovering the true value of the coefficients, recovering the true rankings from data consisting of partial rankings, and predicting the rankings of new alternatives. We compare this method with two different approaches based on the ROL model that include covariates: (i) an approach that does assume heterogeneity between groups and therefore fits $K$ separate models, but does not allow for information sharing, called rank-ordered logit (ROL) (cf.\ Cheng et al., \citeyear{cheng2010label}; Yıldız et al., \citeyear{yildiz2020fast}) and (ii) an approach that does not assume heterogeneity between groups and fits a single model on the pooled data, called pooled rank-ordered logit (PROL). The greater the heterogeneity in preference patterns between groups, the worse the performance of PROL will be. Conversely, ROL is expected to show subpar performance whenever the preference patterns are more homogeneous between groups. 
\\~\\
\noindent The data is simulated in the following manner: for $p \in \{5, 10, 25\}$ we sample a vector of coefficients $\bm{\beta}^{(1)} \sim U(-1,1)$, where group 1 is the baseline group. To induce sparsity, we randomly set $\lfloor\eta p\rfloor$ of the coefficients in $\bm{\beta}^{(1)}$ equal to 0, where $\eta \in \{0.2, 0.8\}$ denotes the (approximate) proportion of sparse coefficients. Subsequently, for $k = 2,\ldots,K$, with $K = 4$, different coefficients are computed by sampling $\lfloor\delta p\rfloor$ coefficients in $\bm{\beta}^{(1)}$ from $U(-1,1)$, where $\delta \in \{0.25, 0.5\}$ denotes the (approximate) proportion of heterogeneity in coefficients compared to the baseline group. The proportion of heterogeneity between subgroups $k$ and $k'$, where $k \neq 1 \neq k' $, is always at least as large as $\delta$. Given the values of $\eta$ and $\delta$, these simulations are referred to the “favourable" simulations, as they reflect mild to strong sparsity and mild to average heterogeneity between groups. The “unfavourable" simulations consist of $\eta \in \{0, 0.2, 0.8\}$ and $\delta \in \{0, 1\}$, which resemble scenarios consisting of no to strong sparsity and either no or complete heterogeneity between groups, and where respectively the PROL and ROL approaches are expected to perform best. The covariates themselves are sampled as $\bm{X} \sim N_p(\bm{0}, \bm{I})$, with identity matrix $\bm{I}$. The true utilities for group $k$ are then obtained from $\exp\left(\bm{\bm{X}\beta}^{(k)}\right)$, whose order reflects the true ranking. To sample the ranking data, the probabilities for the $\binom{M}{m}$ permutations for the partial rankings need to be computed. However, as $M$ grows, for any large $m$, it becomes impossible to compute the $\binom{M}{m}$ permutations, due to the combinatorial explosion that occurs. Therefore, in this simulation study, whenever $p < 25$, we limit ourselves to the scenario whereby individuals rank three out of 20 alternatives, corresponding to situations found in tricot analyses, which provide the application for the proposed method and are further described in Section \ref{Real World Data}, in which individuals only rank three alternatives from a total choice set of $M > 3$ alternatives, where the alternatives are randomly assigned to each individual. However, to ensure identifiability of the $\bm{\beta}^{(k)}$, we set $M = p$ whenever $p = 25$. For each group $k$, the probabilities for all permutations are computed using Equation (\ref{eq:probrank}), which we subsequently use to sample $n_k$ partial rankings, where $n_k \in \{25, 50, 100, 250\}$. To account for the sampling variability of the data, for each combination of parameters, 50 different datasets are generated. 
\\~\\
\noindent With the data created, the models can be fitted. Model performance is evaluated using three different metrics, each serving a different goal. The first of these is the well-known Root Mean Square Error (RMSE): $\sqrt{\frac{\sum_{k=1}^K\sum_{q = 1}^p\left(\beta_q^{(k)} - \hat{\beta}_q^{(k)}\right)^2}{Kp}}$, corresponding to the difference between the estimated and true values for the $\bm{\beta}^{(k)}$. The second metric is the $F_1$ score: $\frac{2\text{tp}}{2\text{tp} + \text{fp} + \text{tn}}$, where tp stands for true positives, fp for false positives and tn for true negatives, and indicates how well the model differentiates between signal $\beta_q^{(k)} \neq 0$ and noise $\beta_q^{(k)} = 0$. The third metric, the Rank Correctness Ratio (RCR): $\frac{1}{K}\sum_{k=1}^K\frac{\sum_{j=1}^M\mathbbm{1}\left(\pi_{j}^{(k)} = \hat{\pi}_{j}^{(k)}\right)}{M}$, is used to evaluate how well the true ranking is recovered by the model. For the RMSE, values closer to 0 indicate better performance, whilst the $F_1$ and RCR scores are indicative of better performance when they attain values closer to 1. 

\begin{figure}[H]
\centering
\text{$M = 20, m = 3, p = 5$}\\
\includegraphics[width=0.25\textwidth]{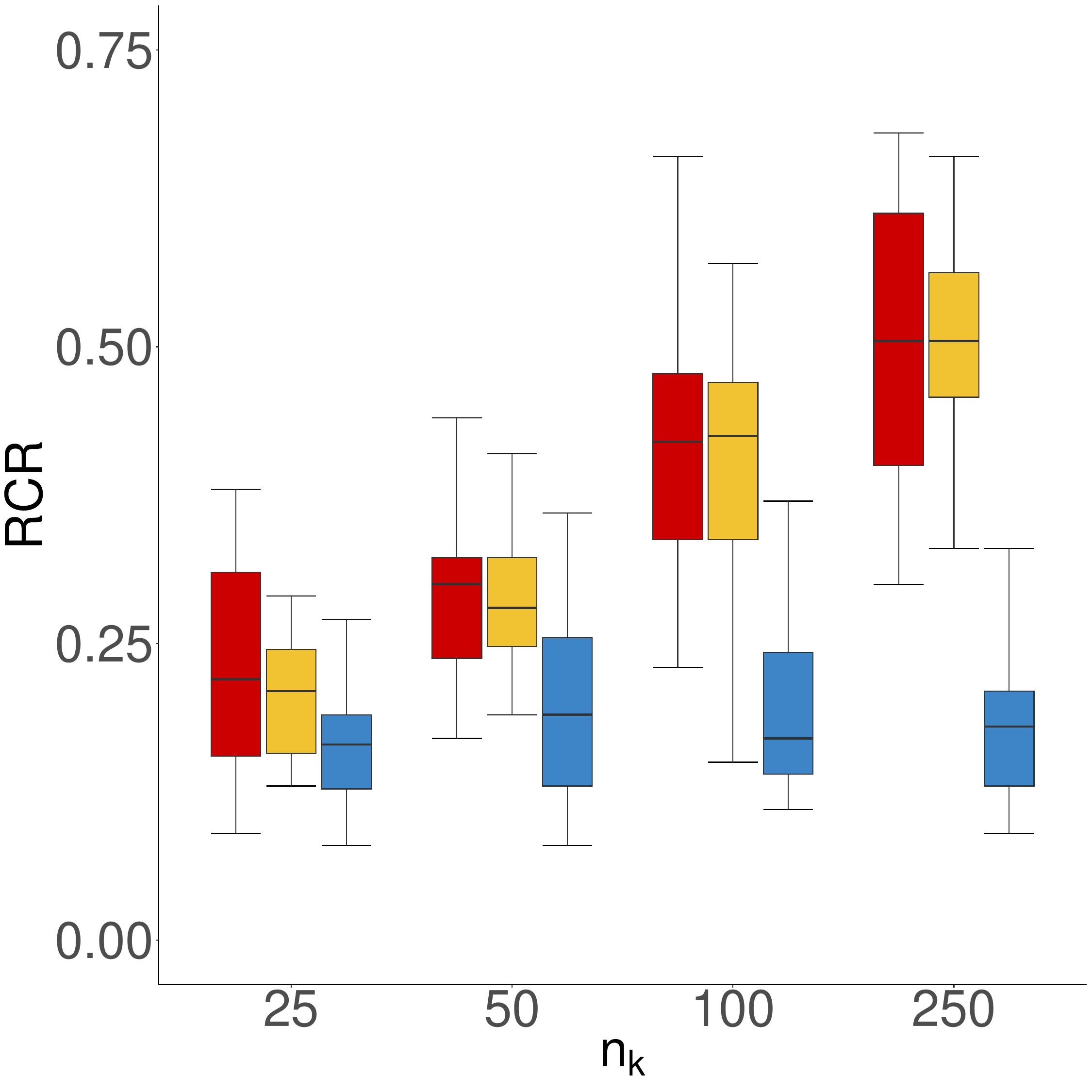}\hfill
\includegraphics[width=0.25\textwidth]{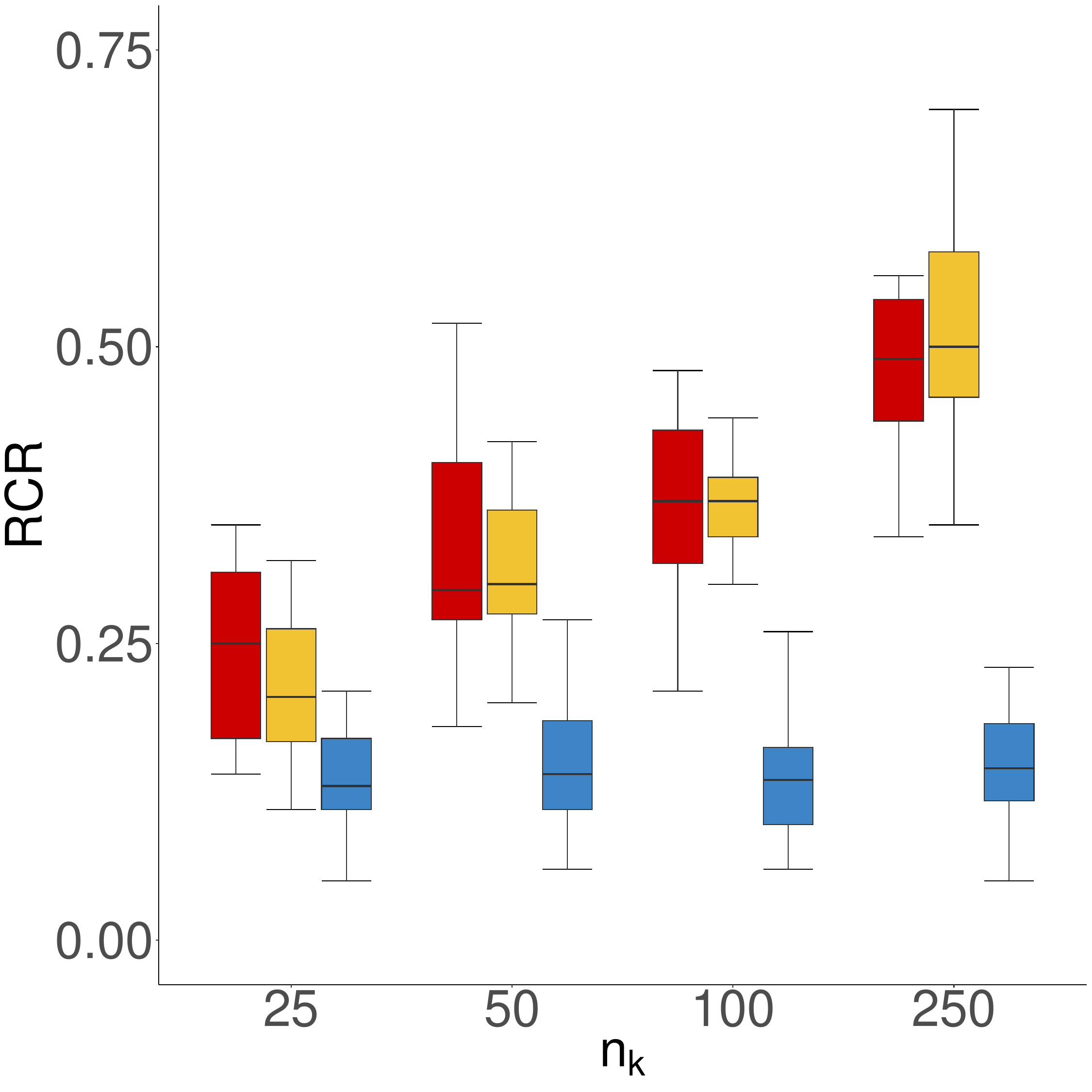}\hfill
\includegraphics[width=0.25\textwidth]{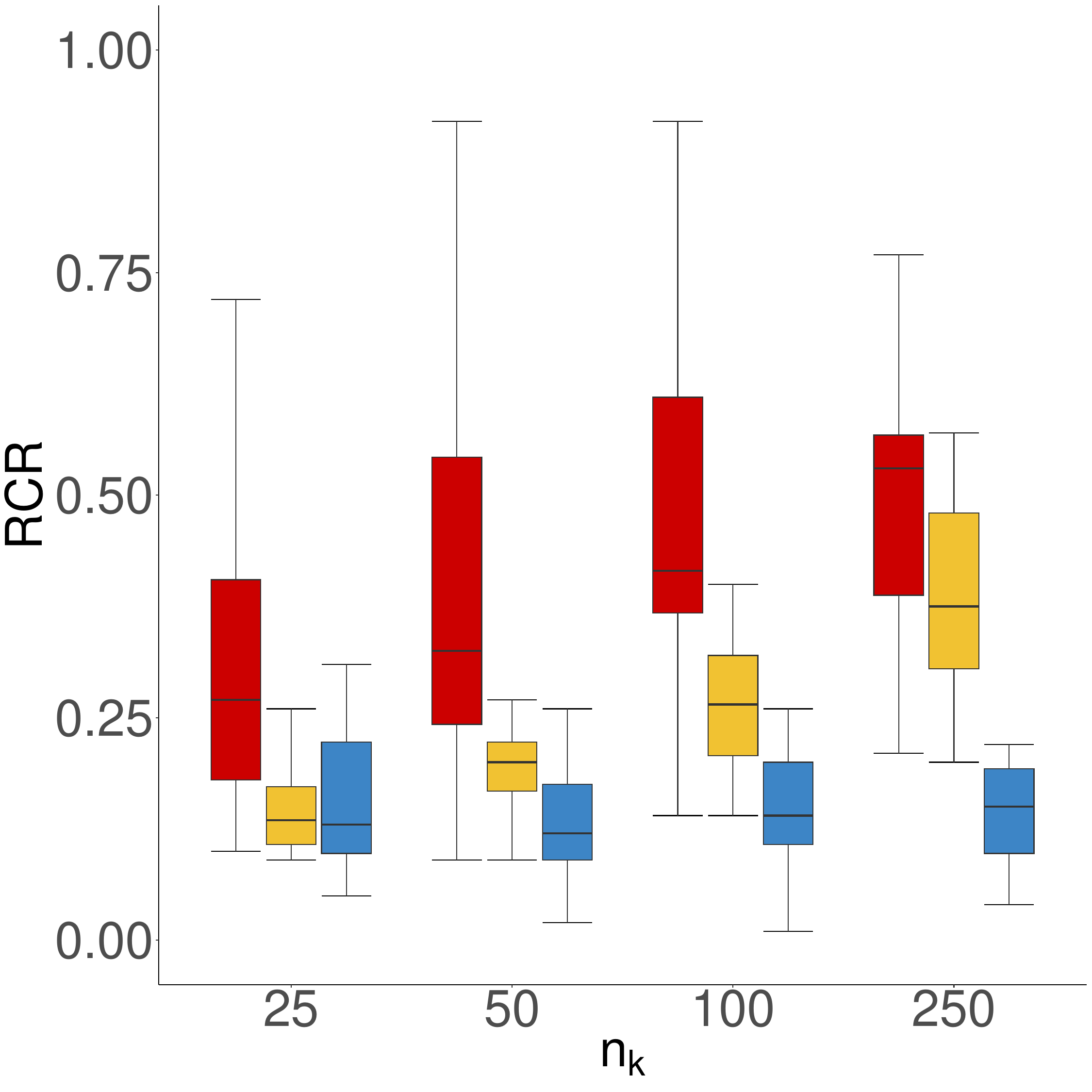}\hfill
\includegraphics[width=0.25\textwidth]{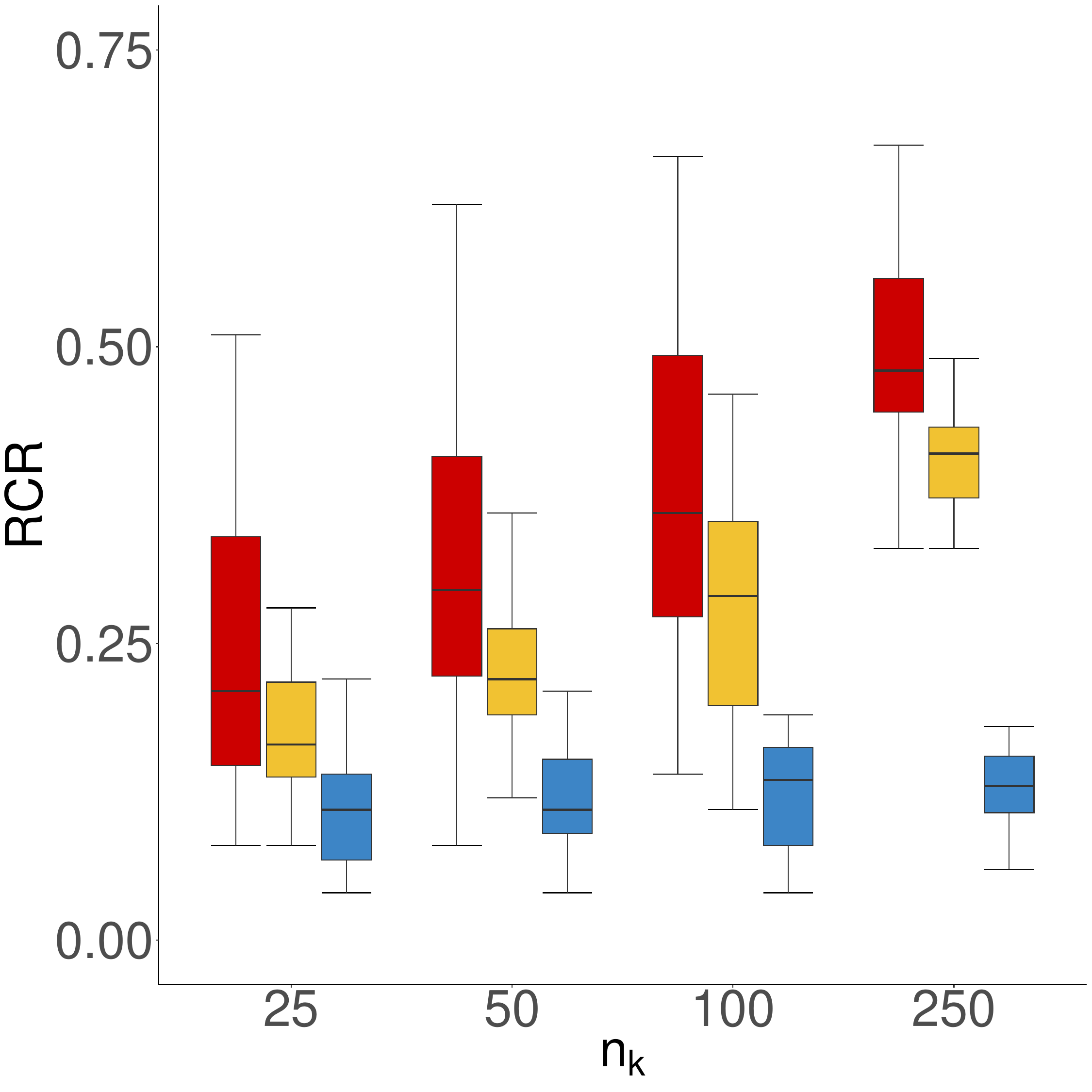}\hfill
\text{$M = 20, m = 3, p = 10$}\\
\includegraphics[width=0.25\textwidth]{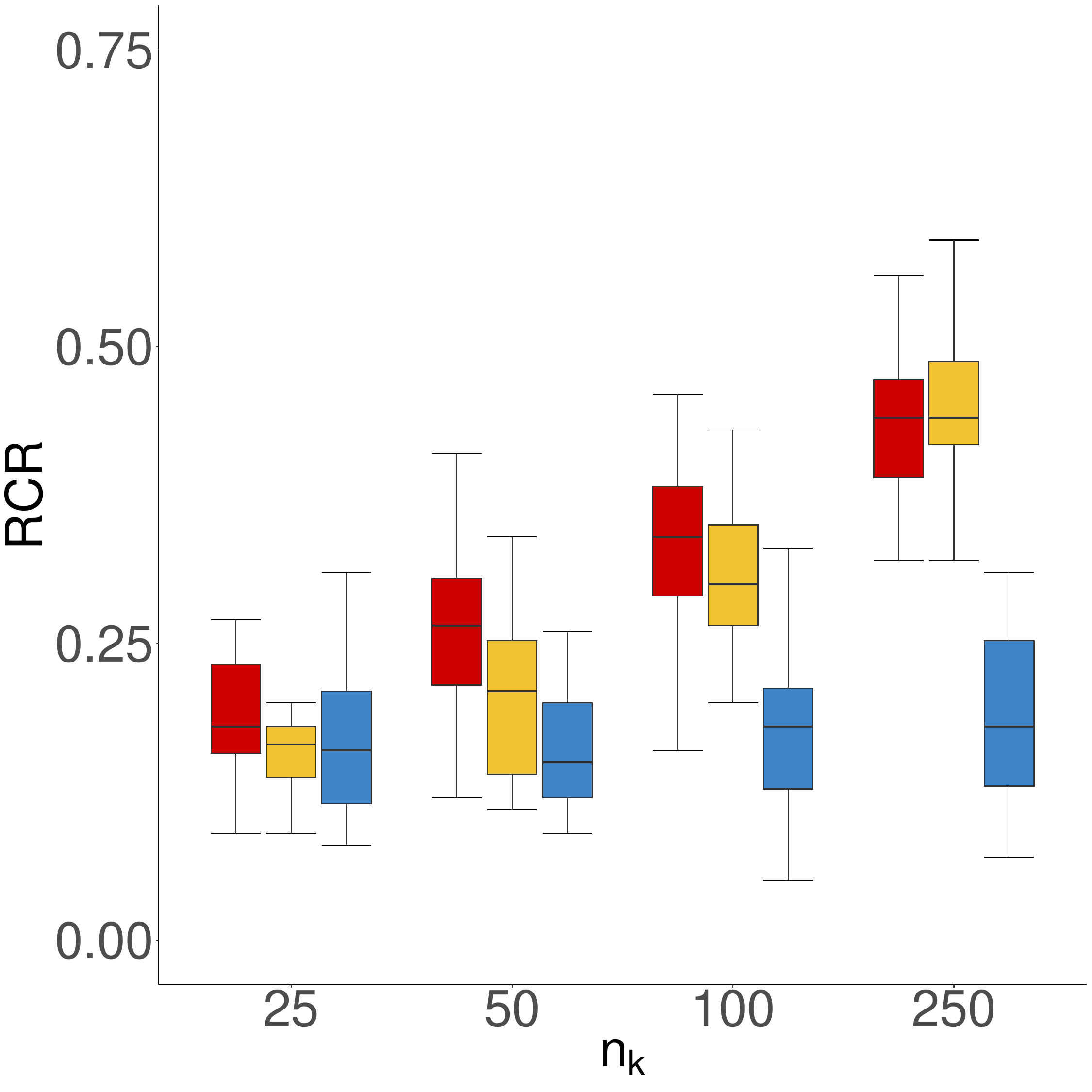}\hfill
\includegraphics[width=0.25\textwidth]{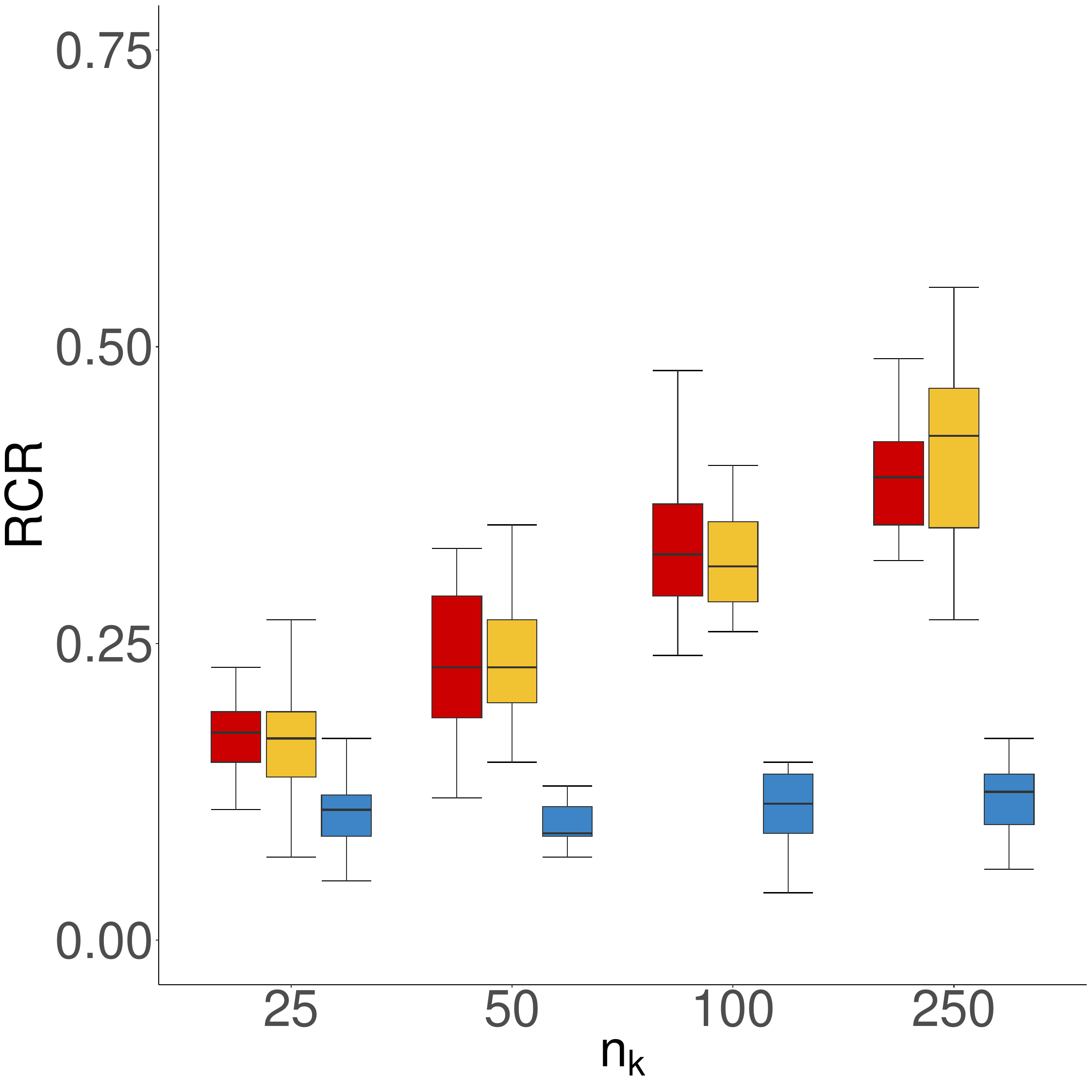}\hfill
\includegraphics[width=0.25\textwidth]{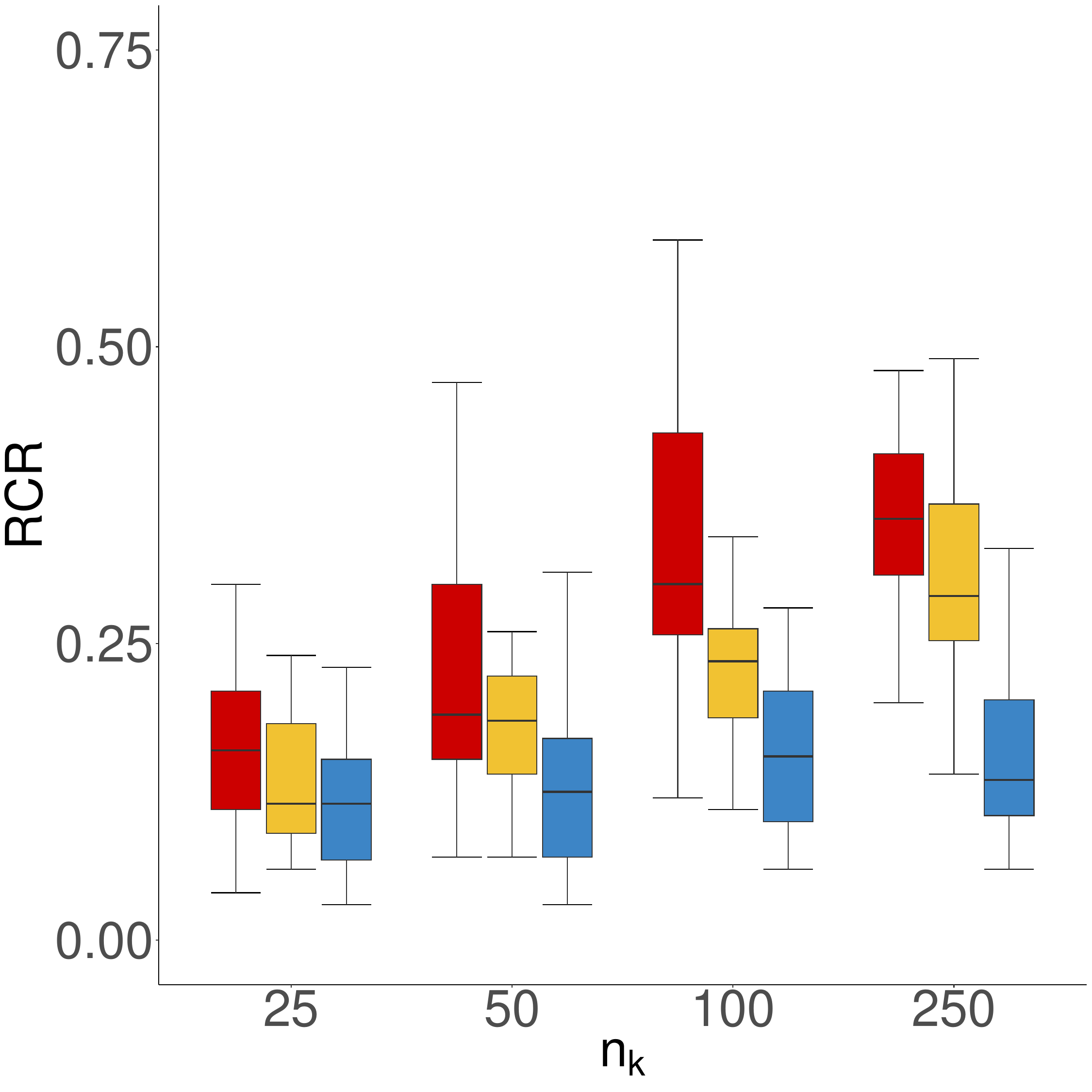}\hfill
\includegraphics[width=0.25\textwidth]{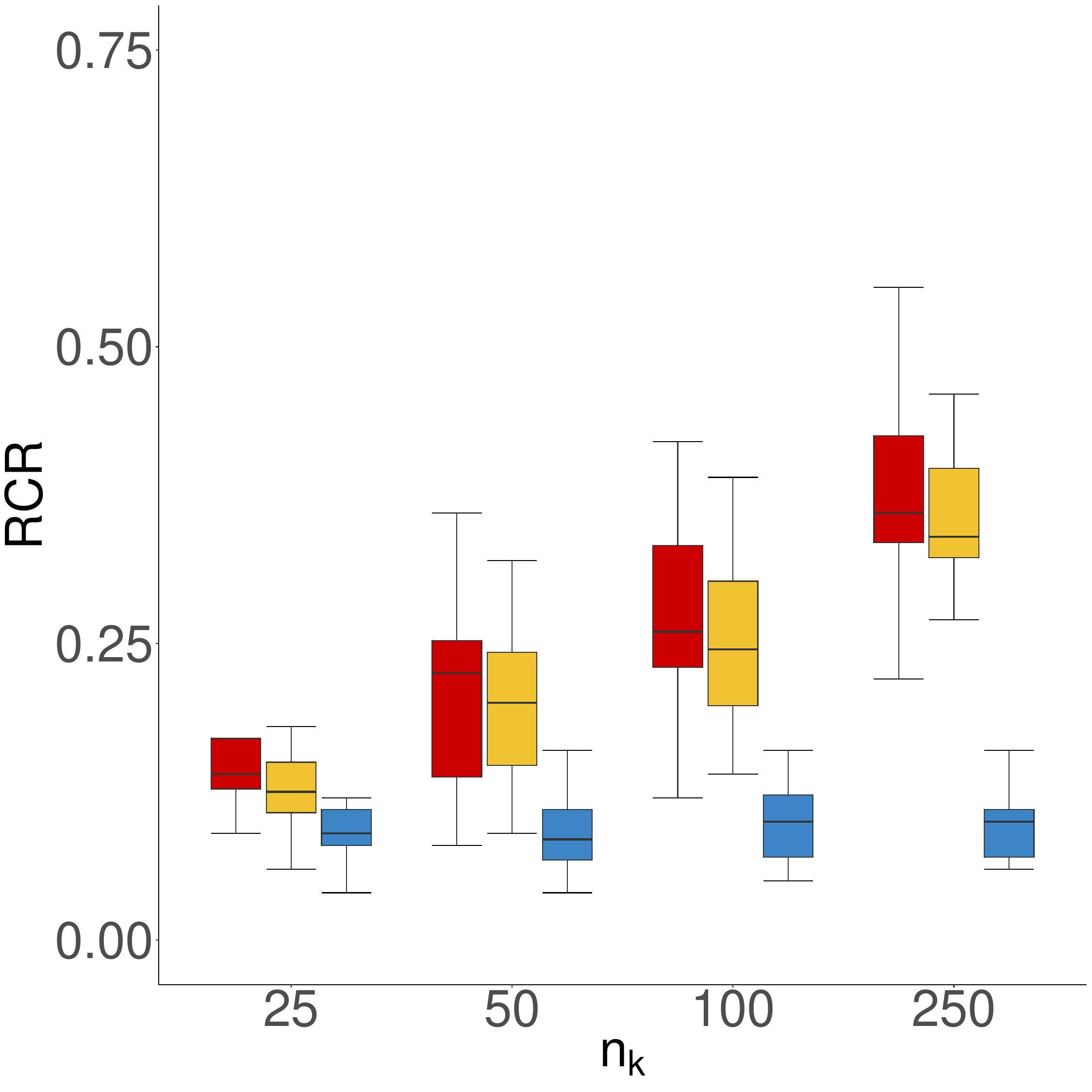}\hfill
\text{$M = 25, m = 3, p = 25$}\\
\includegraphics[width=0.25\textwidth]{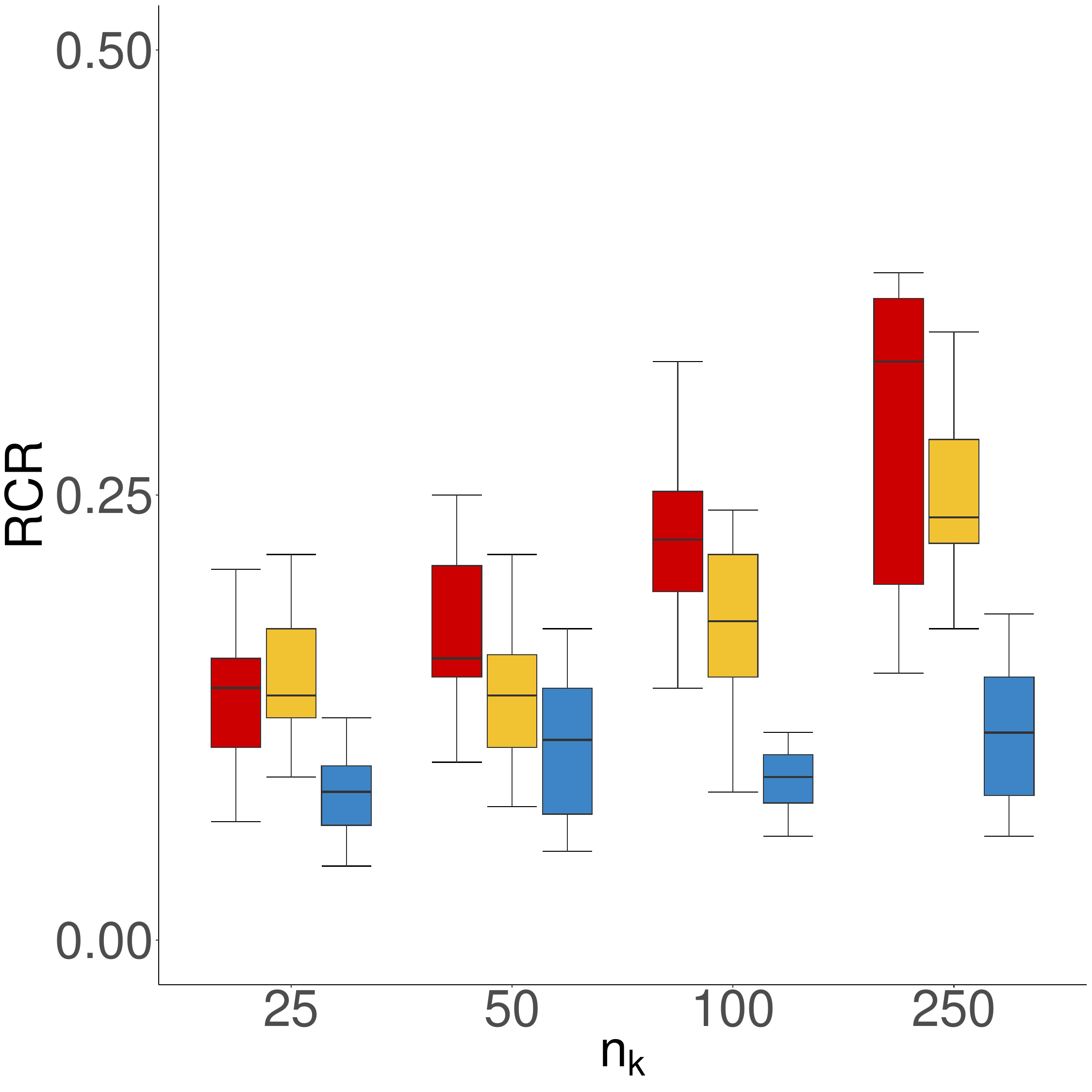}\hfill
\includegraphics[width=0.25\textwidth]{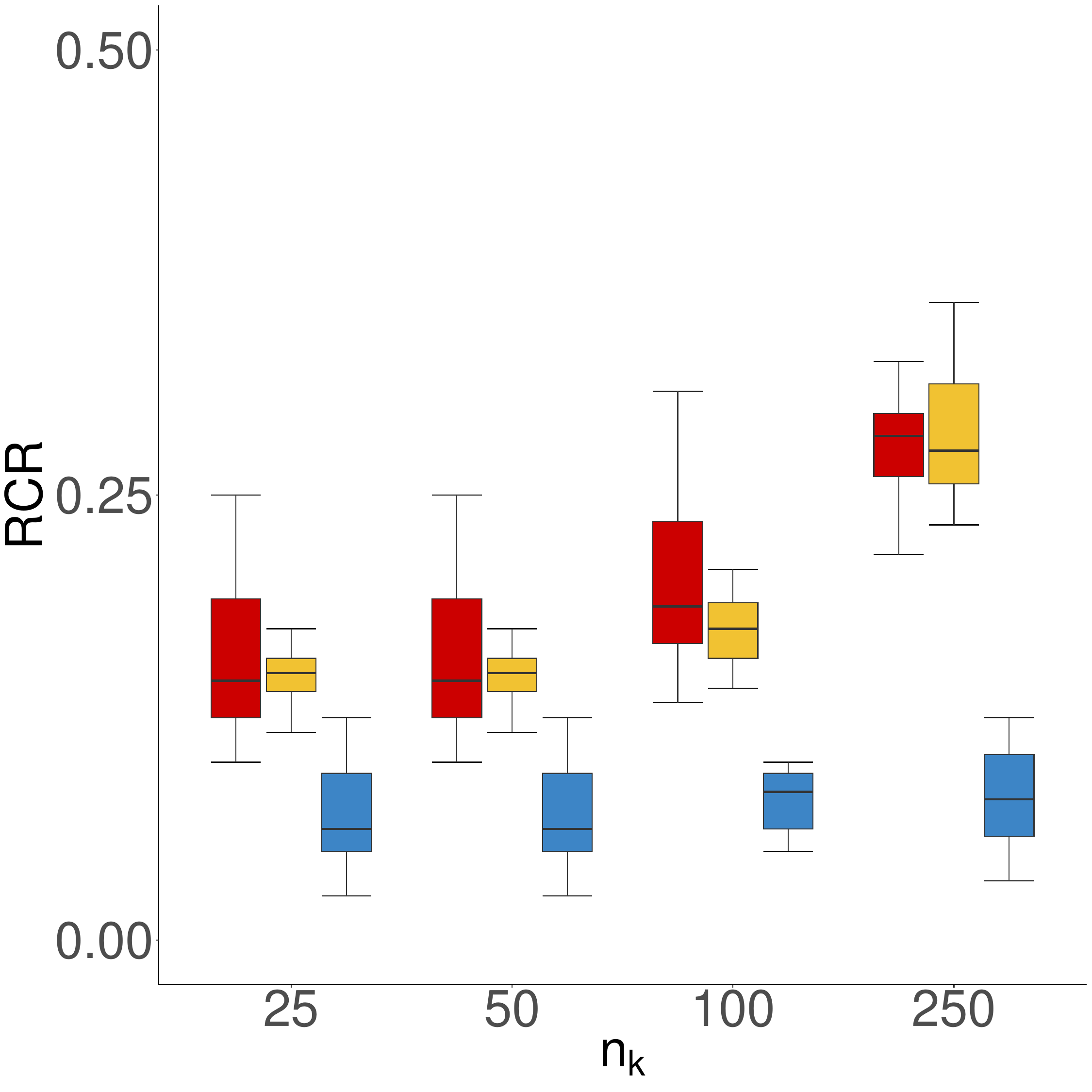}\hfill
\includegraphics[width=0.25\textwidth]{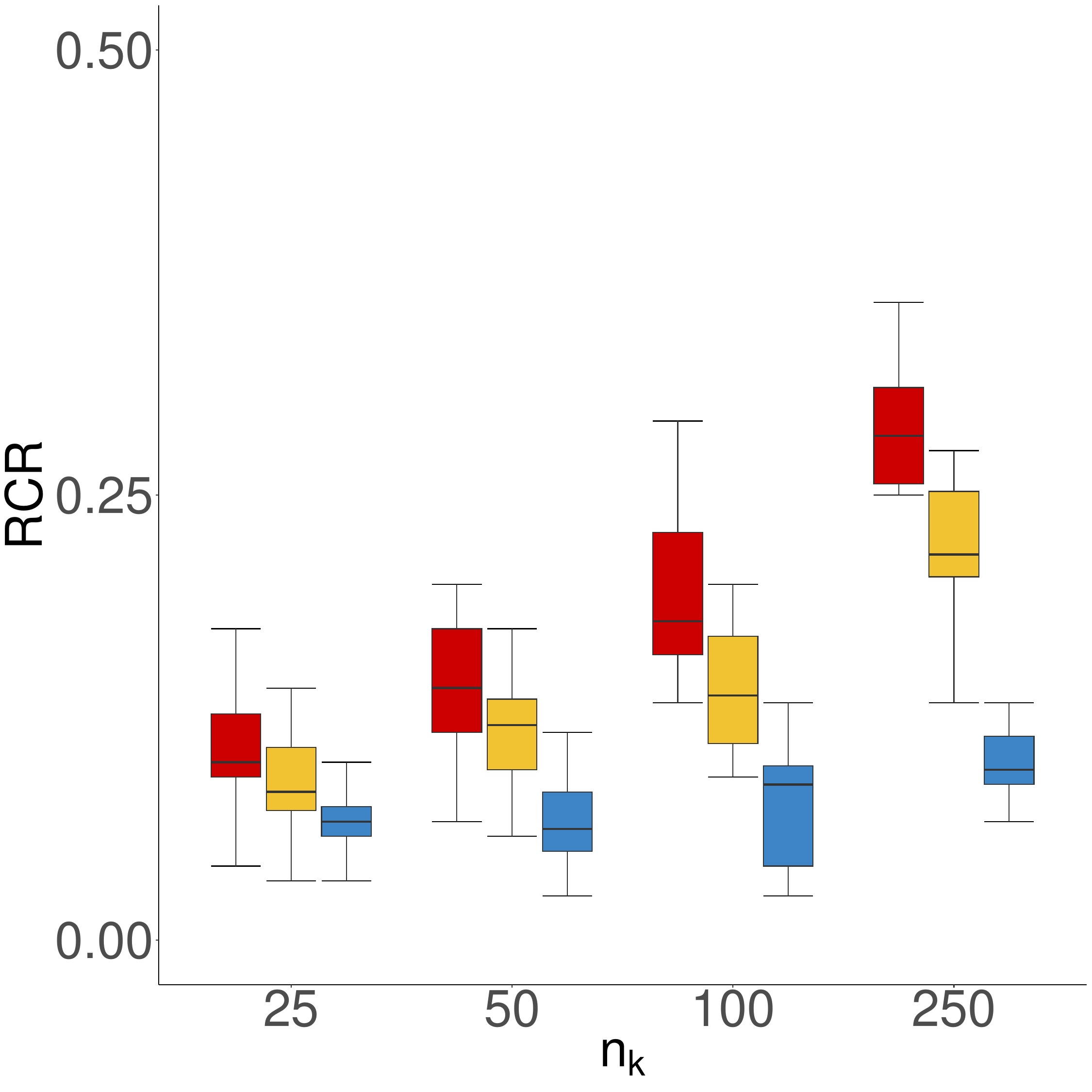}\hfill
\includegraphics[width=0.25\textwidth]{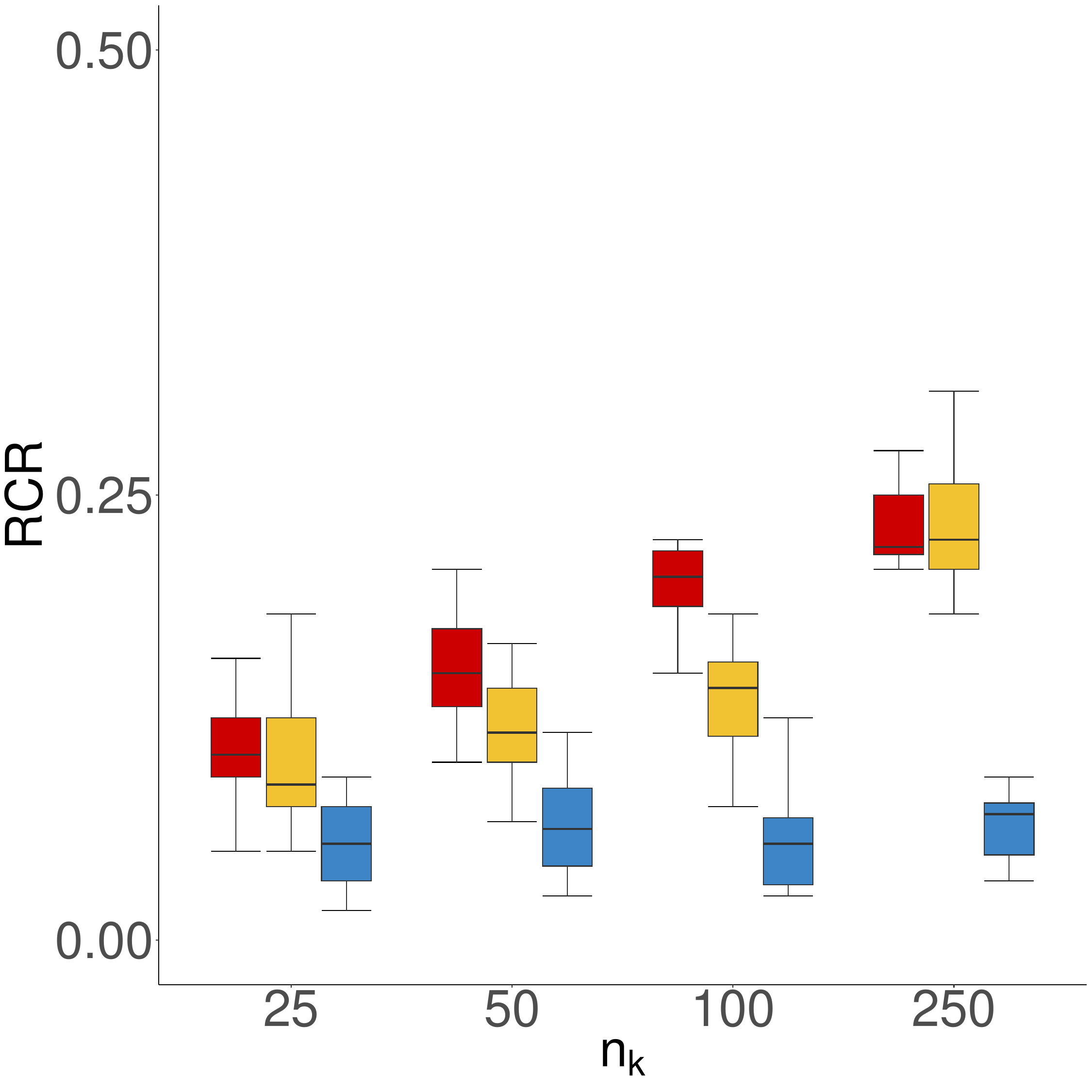}\hfill
\caption[]{Boxplots for the RCR values obtained from 50 different favourable simulated datasets. The colors represent the method used: SFROL \begin{tikzpicture}\draw [fill = colorSFROLAIC] (0,0) rectangle (0.5,0.25);\end{tikzpicture}, ROL \begin{tikzpicture}\draw [fill = colorSFROLBIC] (0,0) rectangle (0.5,0.25);\end{tikzpicture} and PROL \begin{tikzpicture}\draw [fill = colorPROL] (0,0) rectangle (0.5,0.25);\end{tikzpicture}. From left to right, the figures represent the results for $\{\delta = 0.25, \eta = 0.2\}, \{\delta = 0.5, \eta = 0.2\}, \{\delta = 0.25, \eta = 0.8\}$ and $\{\delta = 0.5, \eta = 0.8\}$, where $\delta$ represents the approximate fraction of coefficients that differ across groups, and $\eta$ represents the approximate fraction of sparse coefficients.}
\label{fig:boxplots}
\end{figure}

As prediction is one of the discerning features of the ROL model with covariates, and also the scenario where regularised methods tend to perform best, the simulations also evaluate the model performance on five new alternatives for which only covariates and no observed rankings exist. Accordingly, the RCR is evaluated across $M+5$ alternatives. The results of the favourable simulations for the RCR are shown in Figure \ref{fig:boxplots}, whilst the results for the RMSE and $F_1$ score are shown in Table \ref{tab:simres1}. Results for the RCR of the unfavourable simulations are shown in Table \ref{tab:unfav}. Simulation results for different values of $K$, more informative partial rankings, the unfavourable RMSE and misspecified $K$ are provided in Appendix A in Tables A1, A2, A3 and A4 respectively. 

\begin{table}[H]
\centering
\scalebox{0.60}{%
  \begin{threeparttable}
  \caption{Results for the fitted ROL models on favourable simulated data with $K = 4$, $m = 3$, $M = 20$ if $p < 25$ and $M = p$ otherwise. The RMSE (root mean square error) and $F_1$ scores are averaged across 50 fitted models for each parameter combination and rounded to two decimals. Standard deviations are provided between parentheses. Bold values represent the best result for that particular parameter combination across the three methods. SFROL represents the proposed method, whilst ROL and PROL represent the regular and pooled ROL methods respectively.}
  \label{tab:simres1}
     \begin{tabular}{c | cc | c | cc | c}
        \toprule
        \midrule
         & \multicolumn{3}{c|}{$\delta = 0.25, \eta = 0.2$} & \multicolumn{3}{c}{$\delta = 0.5, \eta = 0.2$}\\ \midrule
         & \multicolumn{2}{c|}{RMSE} & $F_1$ & \multicolumn{2}{c|}{RMSE} & $F_1$ \\ \midrule
         \textbf{$n_k, p$} & \textbf{SFROL} & \textbf{ROL/PROL} & \textbf{SFROL} & \textbf{SFROL} & \textbf{ROL/PROL} & \textbf{SFROL}\\ \midrule
$25, 5$ & \textbf{0.26} (0.07) & 0.37 (0.10)/0.31 (0.09) & 0.90 (0.05) & \textbf{0.26} (0.07) & 0.33 (0.09)/0.38 (0.08) & 0.92 (0.03)\\ 
$50, 5$ & \textbf{0.20} (0.09) & 0.22 (0.08)/0.30 (0.09) & 0.91 (0.03) & \textbf{0.19} (0.05) & 0.21 (0.05)/0.38 (0.09) & 0.91 (0.04)\\ 
$100, 5$ & 0.15 (0.04) & \textbf{0.13} (0.02)/0.29 (0.09) & 0.91 (0.03) & 0.17 (0.05) & \textbf{0.14} (0.02)/0.37 (0.09) & 0.92 (0.04)\\ 
$250, 5$ & 0.12 (0.05) & \textbf{0.09} (0.02)/0.29 (0.09) & 0.92 (0.03) & 0.13 (0.06) & \textbf{0.09} (0.02)/0.37 (0.09) & 0.91 (0.03)\\ 
$25, 10$ & \textbf{0.26} (0.04) & 1.10 (0.63)/0.31 (0.07) & 0.90 (0.04) & \textbf{0.33} (0.09) & 0.95 (0.50)/0.43 (0.06) & 0.90 (0.06)\\ 
$50, 10$ & \textbf{0.22} (0.07) & 0.40 (0.12)/0.29 (0.07) & 0.90 (0.02) & \textbf{0.28} (0.08) & 0.38 (0.09)/0.42 (0.07) & 0.92 (0.03)\\ 
$100, 10$ & \textbf{0.16} (0.04) & 0.23 (0.03)/0.26 (0.06) & 0.90 (0.02) & \textbf{0.20} (0.05) & 0.21 (0.04)/0.41 (0.06) & 0.93 (0.03)\\ 
$250, 10$ & \textbf{0.12} (0.04) & 0.12 (0.02)/0.26 (0.07) & 0.91 (0.02) & 0.14 (0.06) & \textbf{0.12} (0.03)/0.41 (0.06) & 0.93 (0.02)\\ 
$25, 25$ & \textbf{0.50} (0.26) & 1.10 (0.25)/1.23 (0.45) & 0.88 (0.04) & \textbf{0.00} (0.00) & 0.00 (0.00)/0.00 (0.00) & 0.00 (0.00)\\ 
$50, 25$ & \textbf{0.33} (0.07) & 1.13 (0.38)/0.76 (0.73) & 0.89 (0.03) & \textbf{0.45} (0.24) & 1.04 (0.20)/0.70 (0.47) & 0.90 (0.03)
\\ 
$100, 25$ & \textbf{0.31} (0.09) & 0.83 (0.19)/0.64 (0.36) & 0.90 (0.02) & \textbf{0.35} (0.05) & 0.97 (0.22)/0.48 (0.08) & 0.89 (0.02)\\ 
$250, 25$ & \textbf{0.28} (0.11) & 0.42 (0.06)/0.42 (0.06) & 0.91 (0.02) & \textbf{0.34} (0.08) & 0.41 (0.11)/0.68 (0.70) & 0.90 (0.04)
\\ 
        \midrule
          & \multicolumn{3}{c|}{$\delta = 0.25, \eta = 0.8$} & \multicolumn{3}{c}{$\delta = 0.5, \eta = 0.8$}\\ \midrule
         & \multicolumn{2}{c|}{RMSE} & $F_1$ & \multicolumn{2}{c|}{RMSE} & $F_1$ \\ \midrule
         \textbf{$n_k, p$} & \textbf{SFROL} & \textbf{ROL/PROL} & \textbf{SFROL} & \textbf{SFROL} & \textbf{ROL/PROL} & \textbf{SFROL}\\ \midrule
$25, 5$ & \textbf{0.18} (0.05) & 0.34 (0.09)/0.23 (0.07) & 0.56 (0.12) & \textbf{0.23} (0.07) & 0.30 (0.10)/0.30 (0.05) & 0.65 (0.07)\\ 
$50, 5$ & \textbf{0.14} (0.06) & 0.18 (0.04)/0.22 (0.07) & 0.66 (0.15) & \textbf{0.18} (0.05) & 0.18 (0.05)/0.29 (0.05) & 0.67 (0.1)\\ 
$100, 5$ & \textbf{0.11} (0.04) & 0.13 (0.03)/0.21 (0.07) & 0.57 (0.08) & 0.14 (0.04) & \textbf{0.12} (0.02)/0.28 (0.05) & 0.67 (0.08)\\ 
$250, 5$ & 0.09 (0.05) & \textbf{0.07} (0.01)/0.20 (0.07) & 0.58 (0.11) & 0.09 (0.03) & \textbf{0.07} (0.01)/0.28 (0.05) & 0.68 (0.08)\\ 
$25, 10$ & 0.28 (0.18) & 0.59 (0.22)/\textbf{0.25} (0.05) & 0.55 (0.07) & \textbf{0.33} (0.11) & 0.72 (0.28)/0.37 (0.05) & 0.67 (0.04)\\ 
$50, 10$ & \textbf{0.19} (0.06) & 0.29 (0.06)/0.22 (0.05) & 0.55 (0.07) & \textbf{0.24} (0.08) & 0.30 (0.13)/0.34 (0.03) & 0.70 (0.05)\\ 
$100, 10$ & \textbf{0.13} (0.03) & 0.17 (0.03)/0.19 (0.04) & 0.56 (0.06) & \textbf{0.19} (0.05) & 0.21 (0.05)/0.33 (0.04) & 0.67 (0.05)\\ 
$250, 10$ & 0.12 (0.05) & \textbf{0.10} (0.02)/0.19 (0.04) & 0.56 (0.06) & 0.12 (0.03) & \textbf{0.11} (0.02)/0.32 (0.04) & 0.71 (0.05)\\ 
$25, 25$ & \textbf{0.27} (0.06) & 1.40 (0.39)/0.76 (0.47) & 0.54 (0.03) & \textbf{0.46} (0.29) & 1.21 (0.18)/1.22 (1.57) & 0.65 (0.02)\\ 
$50, 25$ & \textbf{0.22} (0.04) & 1.36 (0.31)/0.44 (0.17) & 0.54 (0.03) & \textbf{0.29} (0.04) & 1.19 (0.35)/0.77 (0.83) & 0.66 (0.03)
\\ 
$100, 25$ & \textbf{0.19} (0.06) & 0.69 (0.16)/0.47 (0.24) & 0.56 (0.04) & \textbf{0.35} (0.34) & 0.85 (0.32)/0.41 (0.12) & 0.67 (0.03)
\\ 
$250, 25$ & \textbf{0.12} (0.03) & 0.37 (0.13)/0.38 (0.17) & 0.58 (0.04) & \textbf{0.27} (0.12) & 0.41 (0.12)/0.45 (0.12) & 0.68 (0.03)
\\ 
        \midrule
        \bottomrule
     \end{tabular}
  \end{threeparttable}}
\end{table}

\begin{table}[H]
\centering
\scalebox{0.60}{%
  \begin{threeparttable}
  \caption{Results for the fitted ROL models on unfavourable simulated data with $K = 4$, $m = 3$, $M = 20$ if $p < 25$ and $M = p$ otherwise. The RCR (ranking correctness ratio) is averaged across 50 fitted models for each parameter combination and rounded to two decimals. Standard deviations are provided between parentheses. Bold values represent the best result for that particular parameter combination across the three methods. SFROL represents the proposed method, whilst ROL and PROL represent the regular and pooled ROL methods respectively.}
  \label{tab:unfav}
     \begin{tabular}{c | cc | cc | cc}
        \toprule
        \midrule
         & \multicolumn{2}{c|}{$\delta = 0, \eta = 0$} & \multicolumn{2}{c|}{$\delta = 0, \eta = 0.2$} & \multicolumn{2}{c}{$\delta = 0, \eta = 0.8$}\\ \midrule
         & \multicolumn{2}{c|}{RCR} & \multicolumn{2}{c|}{RCR} & \multicolumn{2}{c|}{RCR} \\ \midrule
         \textbf{$n_k, p$} & \textbf{SFROL} & \textbf{ROL/PROL} & \textbf{SFROL} & \textbf{ROL/PROL} & \textbf{SFROL} & \textbf{ROL/PROL}\\ \midrule
$25, 5$ & 0.29 (0.15) & 0.21 (0.08)/\textbf{0.36} (0.13) & 0.30 (0.16) & 0.21 (0.06)/\textbf{0.36} (0.16) & \textbf{0.40} (0.32) & 0.12 (0.05)/0.21 (0.15)\\ 
$50, 5$ & 0.36 (0.17) & 0.28 (0.08)/\textbf{0.49} (0.17) & 0.34 (0.09) & 0.28 (0.08)/\textbf{0.40} (0.13) & \textbf{0.55} (0.37) & 0.17 (0.08)/0.27 (0.14)\\ 
$100, 5$ & 0.47 (0.17) & 0.36 (0.09)/\textbf{0.56} (0.14) & 0.50 (0.18) & 0.39 (0.08)/\textbf{0.51} (0.18) & \textbf{0.57} (0.31) & 0.20 (0.11)/0.31 (0.17)\\ 
$250, 5$ & 0.58 (0.15) & 0.52 (0.09)/\textbf{0.70} (0.11) & 0.64 (0.15) & 0.50 (0.09)/\textbf{0.66} (0.15) & \textbf{0.60} (0.31) & 0.29 (0.11)/0.44 (0.21)\\ 
$25, 10$ & 0.27 (0.10) & 0.16 (0.04)/\textbf{0.32} (0.08) & 0.25 (0.10) & 0.15 (0.04)/\textbf{0.33} (0.13) & \textbf{0.28} (0.20) & 0.11 (0.05)/0.21 (0.11)\\ 
$50, 10$ & 0.37 (0.13) & 0.23 (0.05)/\textbf{0.42} (0.17) & 0.30 (0.10) & 0.22 (0.07)/\textbf{0.39} (0.17) & \textbf{0.39} (0.20) & 0.16 (0.06)/0.26 (0.11)\\ 
$100, 10$ & 0.48 (0.18) & 0.30 (0.06)/\textbf{0.50} (0.17) & 0.43 (0.12) & 0.32 (0.07)/\textbf{0.51} (0.14) & \textbf{0.50} (0.27) & 0.20 (0.09)/0.35 (0.15)\\ 
$250, 10$ & 0.55 (0.18) & 0.43 (0.10)/\textbf{0.65} (0.14) & 0.51 (0.14) & 0.43 (0.07)/\textbf{0.60} (0.13) & \textbf{0.52} (0.18) & 0.27 (0.11)/0.42 (0.19)\\ 
$25, 25$ & \textbf{0.20} (0.06) & 0.11 (0.03)/0.17 (0.07) & \textbf{0.17} (0.07) & 0.11 (0.03)/0.16 (0.08) & \textbf{0.17} (0.11) & 0.08 (0.04)/0.14 (0.08)\\ 
$50, 25$ & \textbf{0.23} (0.09) & 0.14 (0.03)/0.20 (0.08) & \textbf{0.25} (0.08) & 0.14 (0.04)/0.21 (0.09) & \textbf{0.18} (0.07) & 0.09 (0.04)/0.15 (0.08)\\ 
$100, 25$ & \textbf{0.34} (0.08) & 0.19 (0.05)/0.29 (0.10) & 0.30 (0.08) & 0.19 (0.05)/\textbf{0.31} (0.10) & \textbf{0.26} (0.09) & 0.12 (0.04)/0.16 (0.07)\\ 
$250, 25$ & \textbf{0.40} (0.11) & 0.26 (0.04)/0.40 (0.14) & \textbf{0.39} (0.15) & 0.27 (0.05)/0.33 (0.12) & \textbf{0.35} (0.13) & 0.18 (0.05)/0.28 (0.10)\\ 
       \midrule
         & \multicolumn{2}{c|}{$\delta = 1, \eta = 0$} & \multicolumn{2}{c|}{$\delta = 1, \eta = 0.2$} & \multicolumn{2}{c}{$\delta = 1, \eta = 0.8$}\\ \midrule
         & \multicolumn{2}{c|}{RCR} & \multicolumn{2}{c|}{RCR} & \multicolumn{2}{c|}{RCR} \\ \midrule
         \textbf{$n_k, p$} & \textbf{SFROL} & \textbf{ROL/PROL} & \textbf{SFROL} & \textbf{ROL/PROL} & \textbf{SFROL} & \textbf{ROL/PROL}\\ \midrule
$25, 5$ & 0.21 (0.06) & \textbf{0.22} (0.07)/0.09 (0.03) & 0.22 (0.05) & \textbf{0.25} (0.08)/0.07 (0.03) & \textbf{0.22} (0.08) & 0.20 (0.08)/0.08 (0.04)\\ 
$50, 5$ & 0.29 (0.07) & \textbf{0.29} (0.06)/0.09 (0.04) & 0.29 (0.08) & \textbf{0.30} (0.06)/0.08 (0.03) & \textbf{0.28} (0.08) & 0.25 (0.06)/0.09 (0.03)\\ 
$100, 5$ & 0.35 (0.07) & \textbf{0.37} (0.08)/0.08 (0.03) & 0.37 (0.09) & \textbf{0.38} (0.09)/0.08 (0.03) & \textbf{0.35} (0.11) & 0.33 (0.09)/0.09 (0.04)\\ 
$250, 5$ & 0.47 (0.12) & \textbf{0.50} (0.08)/0.09 (0.04) & 0.45 (0.08) & \textbf{0.49} (0.08)/0.08 (0.03) & 0.38 (0.08) & \textbf{0.43} (0.09)/0.1 (0.03)\\ 
$25, 10$ & 0.16 (0.05) & \textbf{0.18} (0.06)/0.07 (0.02) & \textbf{0.16} (0.04) & 0.16 (0.03)/0.07 (0.02) & \textbf{0.16} (0.04) & 0.16 (0.04)/0.07 (0.03)\\ 
$50, 10$ & 0.21 (0.06) & \textbf{0.24} (0.05)/0.06 (0.03) & 0.21 (0.05) & \textbf{0.23} (0.06)/0.08 (0.03) & 0.21 (0.05) & \textbf{0.22} (0.05)/0.08 (0.03)\\ 
$100, 10$ & 0.31 (0.07) & \textbf{0.32} (0.06)/0.06 (0.03) & 0.31 (0.06) & \textbf{0.32} (0.07)/0.07 (0.03) & 0.26 (0.07) & \textbf{0.29} (0.05)/0.07 (0.03)\\ 
$250, 10$ & 0.41 (0.07) & \textbf{0.45} (0.06)/0.06 (0.03) & 0.40 (0.06) & \textbf{0.43} (0.07)/0.07 (0.03) & 0.38 (0.07) & \textbf{0.40} (0.06)/0.07 (0.03)\\ 
$25, 25$ & 0.11 (0.04) & \textbf{0.12} (0.02)/0.05 (0.02) & 0.12 (0.03) & \textbf{0.12} (0.03)/0.04 (0.02) & 0.08 (0.03) & \textbf{0.10} (0.04)/0.04 (0.02)\\ 
$50, 25$ & 0.15 (0.04) & \textbf{0.16} (0.04)/0.06 (0.02) & \textbf{0.16} (0.04) & 0.15 (0.03)/0.05 (0.02) & \textbf{0.15} (0.04) & 0.14 (0.03)/0.06 (0.02)\\ 
$100, 25$ & \textbf{0.19} (0.04) & 0.19 (0.03)/0.05 (0.02) & \textbf{0.19} (0.05) & 0.19 (0.04)/0.05 (0.02) & \textbf{0.18} (0.04) & 0.18 (0.04)/0.05 (0.02)\\ 
$250, 25$ & 0.27 (0.06) & \textbf{0.28} (0.05)/0.05 (0.02) & \textbf{0.26} (0.06) & 0.27 (0.05)/0.05 (0.02) & \textbf{0.27} (0.04) & 0.25 (0.05)/0.06 (0.02)\\ 
        \midrule
        \bottomrule
     \end{tabular}
  \end{threeparttable}}
\end{table}

\begin{figure}[H]
\centering
\includegraphics[width=0.43\textwidth]{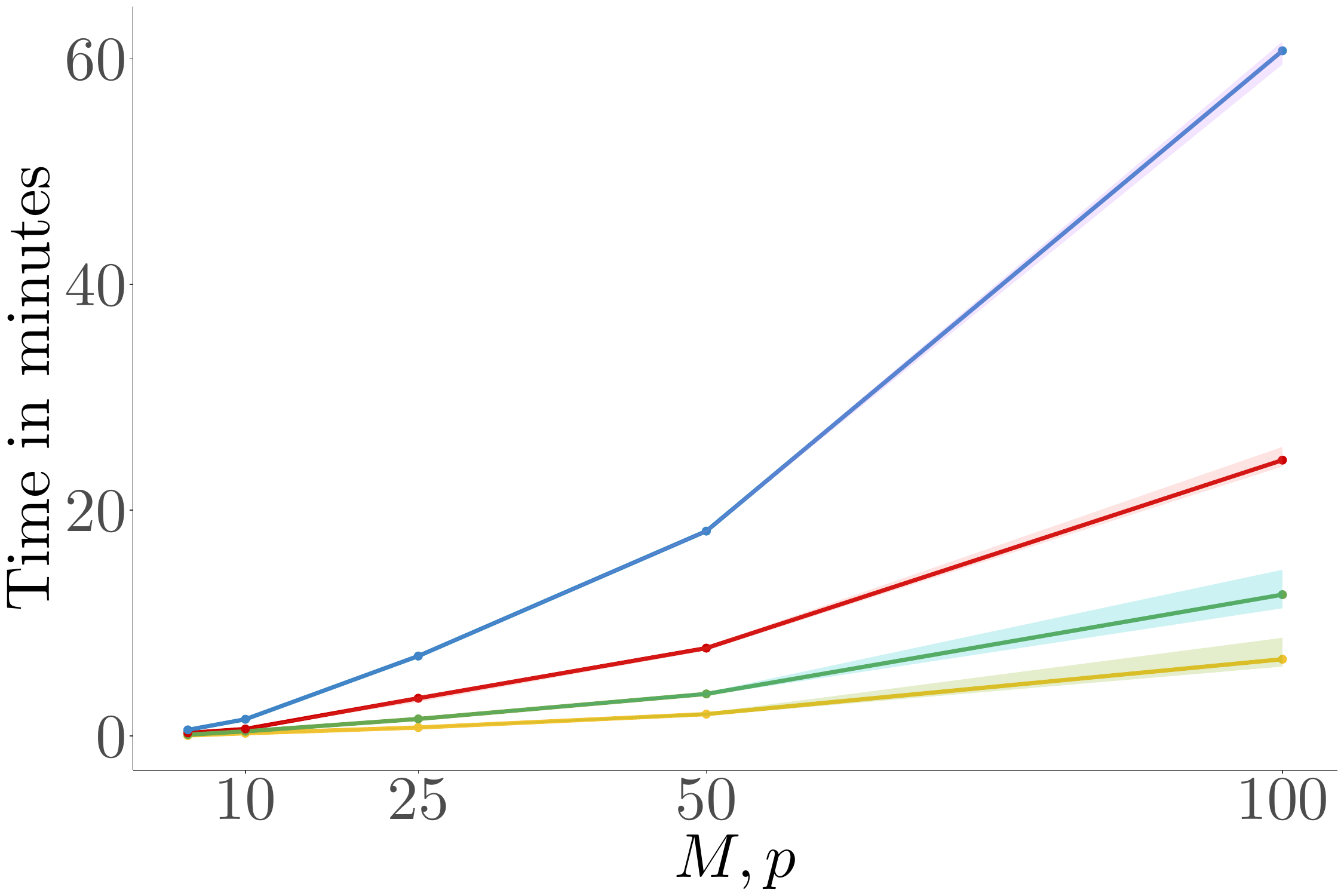}
\caption[]{Computation time for the proposed method in minutes. The colours represent the values of $n_k$: 25 \begin{tikzpicture}\draw [thick, colorSFROLBIC] (0,0.5) -- (0.5,0.5);\end{tikzpicture}, 50 \begin{tikzpicture}\draw [thick, colorROL] (0,0.5) -- (0.5,0.5);\end{tikzpicture}, 100 \begin{tikzpicture}\draw [thick, colorSFROLAIC] (0,0.5) -- (0.5,0.5);\end{tikzpicture} and 250 \begin{tikzpicture}\draw [thick, colorPROL] (0,0.5) -- (0.5,0.5);\end{tikzpicture}. The solid lines reflect the average times across 20 fitted models, whereas the shading around the lines reflect the minimum and maximum computation times across the fitted models.}
\label{fig:timeplot}
\end{figure}

\noindent The results in Table \ref{tab:simres1} show that maximum likelihood estimates are unable to accurately estimate the $\bm{\beta}^{(k)}$ under high-dimensional settings, whilst the proposed method does not have this problem. In fact, the proposed method outperforms the alternatives the vast majority of the favourable simulated settings. Whilst this advantage is by and large also reflected in the estimation of the ranks in Figure \ref{fig:boxplots}, the advantage is diminished whenever the heterogeneity in the coefficients across groups is increased. This makes sense, as the method is less able to use the observations in the other groups to estimate the coefficients and therefore construct rankings based on this borrowed information. Moreover, the fact that the proposed method gains an even bigger comparative edge over the alternatives -- both the RMSE and RCR improve -- whenever the sparsity level is increased is not surprising either, as the other approaches do not impose any penalties for non-sparse coefficients. An interesting secondary observation from these results is that whilst pooling the data across groups can result in estimated coefficients that are close (in terms of RMSE) to those of the proposed method, this is not reflected by the estimated rankings of the pooled approach, which show consistent underperformance. What is more, the pooling of data appears to bias the estimates to the extend that the pooled approach seems unable to learn the rankings from the data. However, when the data generating mechanism is the same for all groups groups ($\delta = 0$), pooling the data results in more accurate point estimation and prediction as judged by Table \ref{tab:unfav}. Similarly, when the groups are completely dissimilar ($\delta = 1$), fitting a ROL model separately on each group tends to work best. However, for large $p$, or when the $\bm{\beta}^{(k)}$ are sparse, the proposed method outperforms the competing approaches, even under unfavourable conditions. Therefore, it appears that the fusion advantages exist primarily for low $n_k$ in relation to $p$. Conversely, the sparsity advantage is more robust for higher values of $n_k$. 
\\~\\
\noindent Next, we showcase the computation time of the proposed method, where the data is simulated in the same way as for the other simulations, except that in this case, we fix $\delta = 0.25$ and $\eta = 0.8$. The results in Figure \ref{fig:timeplot} reflect the average (across 20 different attempts) time in minutes it takes the proposed method to estimate the $\bm{\beta}^{(k)}$ for a single combination of $\lambda_s$ and $\lambda_f$. The simulations were conducted using an AMD Ryzen 5 2600 3.4Ghz processor with 16GB of RAM.
 
\section{Application to preference tricot data} \label{Real World Data}
\noindent We apply the proposed method on what is known as “tricot" -- triadic comparisons of technologies -- data. Originating from agricultural science, tricots are part of the citizen science approach, whereby citizens assist in the collection of crop variety data, such that ultimately, large quantities of preference data are collected. In brief, tricots proceed by researchers handing out three different alternatives to citzens from a larger choice set of $M$ alternatives, where, subsequently, the citizens are asked to rank these alternatives from best to worst on one or multiple traits (van Etten et al., \citeyear{van2019first}; Olaosebikan et al., \citeyear{olaosebikan2024drivers}). Essentially tricot data comprise a subset of best-worst scaling data (Schuster et al., \citeyear{schuster2024rise}). The aim of the researchers is then to construct a full ranking of the $M$ alternatives based on the partial ranking data, which is possible using ROL (van Etten et al., \citeyear{van2019first}). However, the lack of inclusion of covariates in the analyses of tricot data limits the understanding of researchers concerning the determinants of the observed rankings and the ability to predict the rank of new alternatives for covariates exists. As a result of the imposed sparsity penalty, the ensuing variable selection facilitates ease of understanding the ranking process when many covariates are included and causes prediction to be more robust against overfitting. The fusion penalty benefits the analysis of tricot data, as these data are typically gathered under heterogeneous conditions, e.g.\ different locations or different groups of individuals, potentially resulting in biased or highly variable parameter estimates when respectively either too much (pooling the data) or too little (fitting a separate model per group) information across groups is borrowed, provided that the groups share some commonalities. 
\\~\\
\noindent The tricot data used in this application consists of a tasting experiment of sweet potato (\textit{Ipomoea batatas [L.] Lam.}) that Moyo et al.\ (\citeyear{moyo2021consumer}) performed to determine overall preferences for different varieties. Consumers in Ghana and Uganda were approached and asked to partake in the experiment. Upon agreeing, the consumers were provided with three different raw, unpeeled roots of sweet potato varieties. These three varieties were randomly selected from a pool of 21 Ghanaian varieties, consisting of the SARI-Diedi (Tu-Purple), PG17206-N5, CRI-Ligri, PG17265-N1, SARI-Nyumingre (Obare), PG17140-N2, CRI-Apomuden, PG17136-N1, PG17362-N1, CIP442162, SARI-Nan, PG17412-N2, PG17305-N1, PGN16024-27, PGN16021-39, PGN16024-28, PGN16030-30, PGN16092-6, PGN16130-4, PGN16203-18 and PGA14011-24 varieties. The PG-varieties are in the later stages of the breeding process whilst the others are released varieties. The consumers were instructed to prepare the roots following their usual preparation method, eat the roots and rank them based on their overall acceptability.

The data consists of 111 partial rankings, and was supplemented with 11 additional covariates from the SweetPotatoBase (http://sweetpotatobase.org) : the dry matter, sugar, beta-carotene, fructose, glucose, sucrose and maltose contents, and the skin colour darkness, flesh colour darkness, root shape and root size. Each of these covariates represent average scores that were obtained from hundreds of phenotyping trials. Even though some of the covariates were evaluated using ordinal scales in individual phenotyping trials; namely skin colour darkness, flesh colour darkness and root shape, we only have access to the averaged (continuous) values for all covariates. Whilst the consumers were only asked to state their overall preference in the experiment, in the form of a partial ranking, and therefore did not necessarily made a comparative evaluation of the three sweet potatoes on all covariates, we do think that in general, the consumers did take into account these covariates in their evaluation, as they are all sensory attributes. This is further confirmed by some comments made by the consumers, mentioning the taste and look of the sweet potatoes.

For this analysis, we created two groups of individuals: one group consisting of men and one consisting of women, where each group consists of 64 and 47 samples respectively. Not only is this particular grouping consistent with a recent research interest in the field of tricot data (Olaosebikan et al., \citeyear{olaosebikan2024drivers}; Voss et al., \citeyear{voss2023innovative}), based on marketing literature, gender appears to be a key determinant in differing consumer preferences (Moss \& Colman, \citeyear{moss2001choices}; Pirlympou, \citeyear{pirlympou2017critical}; Friedman \& Lowengard, \citeyear{friedmann2019gender}). Not accounting for gender through grouping or by including only those partial rankings provided by either men or women is likely to result in heterogeneity within groups, and as a consequence biased results. In addition, we include only those partial rankings collected in the Nyankpala community in northern Ghana in this analysis to ensure that within groups the observations are i.i.d.\ (Olaosebikan et al., \citeyear{olaosebikan2024drivers}). As such, this application aims to illustrate the differences between men and women from Nyankpala in how they evaluate certain properties of sweet potatoes, and how this, in turn, affects the overall rankings of the potatoes. After standardising the covariates, the first step of this analysis consists of applying the cross-validation approach across a grid of $|\Lambda_s \times \Lambda_f|$ penalty parameters, which is chosen in the same fashion as described in Section \ref{Parameter estimation}, in order to select the appropriate penalty parameters. Subsequently, using these selected parameters, the model is fitted on the data. The estimated coefficients are provided in Table \ref{tab:coefficientsapp1}. As we impose lasso-type penalties on the likelihood to improve point estimation and prediction, statistical inference is highly nontrivial due to the lack of valid confidence intervals (Kyung et al., \citeyear{kyung2010penalized}; Goeman et al., \citeyear{goeman2012l1}). Whilst statistical inference for penalised models is an active field of research, variable selection probabilities can be computed using a (nonparametric) bootstrap.
\begin{table}[H]
\centering
\scalebox{0.7}{
  \begin{threeparttable}
 \caption{Coefficients estimated by the proposed method for the boiled sweet potato data. The selection probabilities based on the bootstrap are provided between brackets. All values are rounded to two decimals.}
  \label{tab:coefficientsapp1}
     \begin{tabular}{|l| c | c | c}
        \toprule
        \midrule
 & \multicolumn{2}{c|}{\textbf{Estimate}}\\ \midrule
\textbf{Covariate} & \textbf{Men} & \textbf{Women}\\ \midrule
Dry matter content & 0.57 (0.99) & -0.01 (0.87)\\
Sugar content     &     0.44 (0.02) &  0.00 (0.87)\\
Beta-carotene content & 0.00 (0.76) & 0.00 (0.82)\\
Fructose content  &    0.64 (1.00) & 0.11 (0.88)\\
Glucose content   &     0.10 (0.77) & -0.38 (0.94) \\
Sucrose content   &    0.06 (0.89) & 0.06 (0.88)\\
Maltose content    &   0.00 (0.85) & -0.11 (0.92)\\
Skin colour darkness & -0.27 (0.96) & -0.27 (0.97)\\
Flesh colour darkness & 0.01 (0.78) & 0.00 (0.72)\\
Root shape     &       -0.24 (0.94) & 0.01 (0.88)\\
Root size        &      0.04 (0.91) &  0.21 (0.94)\\
        \midrule
	\bottomrule
     \end{tabular}
 \end{threeparttable}}
\end{table} 

\noindent The estimates provided in Table \ref{tab:coefficientsapp1} are for the most part non-sparse; only one out of the 12 coefficients was estimated to be zero for both men and women. Moreover, there seems to be a substantial amount of similarity between men and women in the relationship that the covariates have on the observed rankings, due to the information sharing between groups as is reflected by our selected penalties $\lambda_s = 0.01$ and $\lambda_f = 0.005$. Nevertheless, a substantial amount of the coefficients differ between the two groups, including three differences between the signs of the estimates.

In contrast to the Nyankpala women who seem to exhibit a preference for less dry sweet potatoes that is consistent with the literature (Martin \& Rodriguez-Sosa, \citeyear{martin1985preference}; Moyo et al., \citeyear{moyo2021consumer}), the fact that the Nyankpala men strongly prefer dry sweet potatoes is remarkable. Conversely, the Nyankpala men do exhibit the more typical preference for the sweeter variants of the sweet potatoes, as is typically observed in the literature (Martin \& Rodriguez-Sosa, \citeyear{martin1985preference}; Leksrisompong et al., \citeyear{leksrisompong2012sensory}; Moyo et al., \citeyear{moyo2021consumer}). In addition to the sugar content, the data consists of measurements on four different kinds of sugar: fructose, glucose, sucrose and maltose. Nevertheless, little is known about preferences with respect to the kinds of sugar, except that the positive effect of glucose for the Nyankpala men on the sweet potato preference might be due to potatoes with a high glucose content tasting like brown sugar (Leksrisompong et al., \citeyear{leksrisompong2012sensory}). Even though the covariates related to the taste of the sweet potato are not always consistent with the existing literature, those related to the colour of the sweet potato are. Preferences for cream or orange coloured sweet potatoes over dark purple or brown ones are well-supported by the literature (Martin \& Rodriguez-Sosa, \citeyear{martin1985preference}), corroborating the negative sign for the flesh colour darkness (ranking from light to dark) found in both Nyankpala men and women. There is little existing research on shape preferences for sweet potato, although a recent study found preferences for more round root shapes (Afuape et al., \citeyear{afuape2021farmers}), corresponding with the preferences of the Nyankpala women, but contradicting the negative sign obtained in the present analysis for the Nyankpala men, as this covariate ranges from round to long shapes. In addition, larger root sizes are preferred. This is not only a recurring preference in the literature (Kapinga et al., \citeyear{kapinga2003farmer}; Afuape et al., \citeyear{afuape2021farmers}; Ahoudou et al., \citeyear{ahoudou2023farmers}), but it can also be deduced from the fact that Ghana has a substantial amount of people who do not have access to enough food (Darfour \& Rosentrater, \citeyear{darfour2016agriculture}), and larger sweet potatoes can feed more mouths. What is remarkable is that this expected positive effect of root size on the preference of individuals holds more strongly for Nyankpala women than for Nyankpala men, who are almost indifferent to the root size. 
\\~\\
\noindent Even though no other sweet potatoes were ranked by the participants of this study, the sweetpotato catalogue for sub-Saharan Africa (Musembi et al., \citeyear{sweetpotato}) provides ample information for other sweet potato varieties. We introduce five other Ghanaian varieties here: Blue-Blue, Faara, CRI-Santom Pona, CRI-Okumkom and CRI-Patron. Covariate information not included in the catalogue was supplemented with data from the SweetPotatoBase (http://sweetpotatobase.org). The predicted ranks for these new varieties are included with the estimated ranks and ordered into an aggregated ranking list. This list is provided in Table \ref{tab:rankingssapp1}. To evaluate the predictive accuracy of the fitted model, we apply a 5-fold cross-validation procedure, where across each fold, the fitted model is used to predict the ratings of the test data. This results in a RCR of 0.53, implying that just over half of the predicted rankings are correct.
\begin{table}[H]
\centering
\scalebox{0.58}{
  \begin{threeparttable}
 \caption{Rankings estimated by the proposed method for the boiled sweet potato data consisting of both the alternatives for which ranking data are available and the alternatives for which there are not.}
  \label{tab:rankingssapp1}
     \begin{tabular}{|l| c | c | c}
        \toprule
        \midrule
\textbf{Variety} & \textbf{Men} & \textbf{Women}\\ \midrule
SARI-Diedi (Tu-Purple)           &  25  & 23 \\			
PG17206-N5        &   7 & 6  \\			
CRI-Ligri     &  22    & 5 \\			
PG17265-N1            &  6   & 24 \\			
SARI-Nyumingre (Obare)     &      13     & 7 \\			
PG17140-N2       &   12      &  4 \\		
CRI-Apomuden    &   2 & 18 \\	
PG17136-N1     &     3  & 9 \\		
PG17362-N1            &   17  & 11 \\			
CIP442162     &      18    &  10 \\	
SARI-Nan       &      11    & 17 \\		
PG17412-N2 & 1  & 14 \\
PG17305-N1 & 15 & 12 \\
PGN16024-27 & 21 & 21 \\
PGN16021-39 & 16 & 26 \\
PGN16024-28 & 24 & 22 \\
PGN16030-30 & 9 & 20 \\
PGN16092-6 & 23 & 15\\
PGN16130-4 & 26 & 25\\
PGN16203-18 & 8 & 19\\
PGA14011-24 & 4 & 8\\
Blue-Blue & 14 & 1\\
Faara & 20 & 13\\
CRI-Santom Pona & 19 & 3\\
CRI-Okumkom & 5 & 16\\
CRI-Patron & 10 & 2\\
\midrule
\bottomrule
     \end{tabular}
 \end{threeparttable}}
\end{table}

\noindent Given that the differences in the estimated coefficients between men and women are nontrivial, with two differing signs, the aggregated ranking lists provided in Table \ref{tab:rankingssapp1} are also quite different as well. Nevertheless, the mediocre value of the RCR (0.53) as obtained by the cross-validation procedure should be kept in mind when discussing these results. What is remarkable about this aggregated ranking list is most of the varieties in the later stages of the breeding process -- the PG-varieties -- are ranked at the bottom of the list. This in contrast to the well-established varieties that are preferred by the individuals. The low rankings for most of the PG-varieties can be explained by their high (low) values on properties deemed undesirable (desirable) by the {individuals partaking in the experiment. 

\section{Conclusion and discussion} \label{Conclusion}
\noindent In this contribution, we introduced a doubly penalised version of the ROL model with covariates; the SFROL model, in order to improve point estimation, prediction and interpretation for sparse multi-group data. The model outperforms existing methods across a large variety of different datasets, illustrating the tangible benefits of the proposed method. Applying the model on tricot data -- a new citizen science based approach to generate large quantities of ranking data -- provides new insights into sweet potato preferences, which breeding companies can utilise to create a better match between new potato varieties and the desires of individuals. 

Whilst the proposed method shares some similarities with the mixture ROL model (Gormley \& Murphy, \citeyear{gormley2008mixture}), our method assumes a known group structure, perhaps motivated by substantive knowledge, utilises covariates and provides penalisation on the coefficients between and within groups. Nevertheless, the proposed method can be extended to detect groups in the data.

On another note, the fusion framework can be applied on the ROL model without covariates. However, this model does not allow for predictions, and as such was not evaluated given the importance attached to prediction in this paper.

For future research, the proposed method can be extended by allowing for covariates corresponding to properties of the individuals, although this would complicate prediction. Another potential extension consists of the usage of mixture models to identify groups of individuals with similar preference patterns, when these groups are unknown a-priori. As the existing literature on theoretical analyses of the asymptotic properties of the ROL model under different settings is very limited, this provides another avenue of future research. Finally, the development of statistical ranking models that are better able to deal with scenarios akin to the tricot scenario $m = 3 \ll M$, is another promising direction, as due to the ease and cost-effectiveness of gathering tricot data, this type of data is expected to be more common in the future. 

\setcounter{figure}{0}
\renewcommand{\thefigure}{A\arabic{figure}}
\setcounter{table}{0}
\renewcommand{\thetable}{A\arabic{table}}

\appendix 
\section*{Appendix A: Additional simulation results} \label{Additional simulation results}
\noindent To showcase that the proposed method performs well under a myriad of different settings, some additional simulation results are shown in the tables below. The data are generated in the same manner as in Section 4. Note that for the simulations in Table \ref{tab:diffm} we restrict ourselves to $p \leq 10$ due to the combinatorial explosion that occurs otherwise when simulation informative rankings.

The first set of additional simulations are those conducted with different values of $K$, namely 2 and 6 instead of 4 used in Section 4. Fixing $\delta$ to 0.25 and $\eta$ to 0.8, the results are provided in Table \ref{tab:diffk}.
\begin{table}[H]
\centering
\resizebox{0.6\textwidth}{!}{%
  \begin{threeparttable}
  \caption{Results for the fitted ROL models on simulated data with $\delta = 0.25, \eta = 0.8$, $M = 20$ and $m = 3$. The RMSE (root mean square error), $F_1$ and RCR (rank correctness ratio) scores are averaged across 50 fitted models for each parameter combination and rounded to two decimals. The same holds for the RCR pred, which indicated the RCR on the aggregated list of the 20 alternatives with five new alternatives, whose rankings are predicted. Standard errors are provided between parentheses. Bold values represent the best result for that particular parameter combination across the three methods. SFROL represents the proposed method, whilst ROL and PROL represent the regular and pooled ROL methods respectively. Both metrics are provided for the SFROL, whilst only the RMSE is provided for the ROL and PROL methods, as these do not allow for variable selection}
  \label{tab:diffk}
     \begin{tabular}{c | cc | c | cc}
        \toprule
        \midrule
         & \multicolumn{5}{c|}{$K = 2$}\\ \midrule
         & \multicolumn{2}{c|}{RMSE} & $F_1$ & \multicolumn{2}{c|}{RCR}\\ \midrule
         \textbf{$n_k, p$} & \textbf{SFROL} & \textbf{ROL/PROL} & \textbf{SFROL} & \textbf{SFROL} & \textbf{ROL/PROL}\\ \midrule
$25, 5$ & \textbf{0.21} (0.08) & 0.27 (0.09)/0.22 (0.08) & 0.43 (0.08) & \textbf{0.18} (0.06) & 0.11 (0.17)/0.13 (0.08)\\ 
$50, 5$ & \textbf{0.12} (0.05) & 0.17 (0.05)/0.18 (0.08) & 0.46 (0.09) & \textbf{0.36} (0.19) & 0.20 (0.09)/0.18 (0.08)\\ 
$100, 5$ & \textbf{0.10} (0.04) & 0.12 (0.03)/0.16 (0.07) & 0.48 (0.10) & \textbf{0.43} (0.23) & 0.24 (0.10)/0.17 (0.10)\\ 
$250, 5$ & \textbf{0.07} (0.02) & 0.08 (0.02)/0.14 (0.08) & 0.50 (0.12) & \textbf{0.49} (0.21) & 0.32 (0.14)/0.22 (0.14)\\ 
$25, 10$ & \textbf{0.25} (0.13) & 0.64 (0.42)/0.32 (0.12) & 0.47 (0.05) & \textbf{0.16} (0.08) & 0.11 (0.06)/0.13 (0.07)\\ 
$50, 10$ & \textbf{0.17} (0.08) & 0.28 (0.07)/0.22 (0.06) & 0.50 (0.08) & \textbf{0.24} (0.15) & 0.13 (0.07)/0.16 (0.07)\\ 
$100, 10$ & \textbf{0.12} (0.04) & 0.17 (0.05)/0.16 (0.05) & 0.48 (0.07) & \textbf{0.30} (0.16) & 0.23 (0.10)/0.21 (0.12)\\ 
$250, 10$ & \textbf{0.10} (0.03) & 0.11 (0.03)/0.14 (0.06) & 0.55 (0.11) & \textbf{0.42} (0.20) & 0.33 (0.14)/0.22 (0.13)\\ 
$25, 25$ & \textbf{0.38} (0.30) & 1.57 (0.65)/1.53 (1.46) & 0.49 (0.04) & \textbf{0.12} (0.04) & 0.09 (0.04)/0.06 (0.03)\\ 
$50, 25$ & \textbf{0.27} (0.29) & 1.53 (0.45)/0.74 (0.69) & 0.52 (0.05) & \textbf{0.17} (0.06) & 0.09 (0.05)/0.09 (0.04)\\ 
$100, 25$ & \textbf{0.26} (0.37) & 0.89 (0.28)/0.45 (0.11) & 0.51 (0.04) & \textbf{0.23} (0.11) & 0.12 (0.04)/0.10 (0.06)\\ 
$250, 25$ & \textbf{0.12} (0.03) & 0.38 (0.11)/0.27 (0.06) & 0.55 (0.05) & \textbf{0.29} (0.11) & 0.21 (0.05)/0.12 (0.07)\\ 
        \midrule
         & \multicolumn{5}{c|}{$K = 6$}\\ \midrule
         & \multicolumn{2}{c|}{RMSE} & $F_1$ & \multicolumn{2}{c|}{RCR}\\ \midrule
         \textbf{$n_k, p$} & \textbf{SFROL} & \textbf{ROL/PROL} & \textbf{SFROL} & \textbf{SFROL} & \textbf{ROL/PROL}\\ \midrule
$25, 5$ & \textbf{0.19} (0.05) & 0.28 (0.05)/0.25 (0.06) & 0.55 (0.07) & \textbf{0.25} (0.15) & 0.16 (0.05)/0.14 (0.07)\\ 
$50, 5$ & \textbf{0.17} (0.04) & 0.19 (0.03)/0.25 (0.06) & 0.57 (0.12) & \textbf{0.28} (0.12) & 0.20 (0.07)/0.14 (0.05)\\ 
$100, 5$ & 0.14 (0.05) & \textbf{0.12} (0.02)/0.24 (0.06) & 0.58 (0.12) & \textbf{0.38} (0.15) & 0.27 (0.07)/0.14 (0.06)\\ 
$250, 5$ & 0.10 (0.05) & \textbf{0.07} (0.01)/0.24 (0.06) & 0.59 (0.11) & \textbf{0.50} (0.14) & 0.38 (0.09)/0.14 (0.06)\\ 
$25, 10$ & \textbf{0.22} (0.06) & 0.68 (0.26)/0.24 (0.04) & 0.54 (0.06) & \textbf{0.16} (0.06) & 0.12 (0.04)/0.12 (0.04)\\ 
$50, 10$ & \textbf{0.18} (0.05) & 0.31 (0.09)/0.23 (0.04) & 0.56 (0.06) & \textbf{0.23} (0.08) & 0.16 (0.03)/0.13 (0.05)\\ 
$100, 10$ & \textbf{0.15} (0.05) & 0.19 (0.05)/0.21 (0.04) & 0.61 (0.08) & \textbf{0.34} (0.11) & 0.25 (0.06)/0.13 (0.05)\\ 
$250, 10$ & \textbf{0.11} (0.03) & 0.11 (0.01)/0.21 (0.04) & 0.57 (0.05) & \textbf{0.36} (0.10) & 0.30 (0.04)/0.14 (0.06)\\ 
$25, 25$ & \textbf{0.39} (0.34) & 1.20 (0.23)/0.53 (0.29) & 0.54 (0.02) & \textbf{0.12} (0.02) & 0.09 (0.02)/0.07 (0.02)\\ 
$50, 25$ & \textbf{0.23} (0.02) & 1.18 (0.24)/0.39 (0.06) & 0.56 (0.03) & \textbf{0.16} (0.04) & 0.11 (0.02)/0.07 (0.02)\\ 
$100, 25$ & \textbf{0.19} (0.02) & 0.65 (0.12)/0.34 (0.07) & 0.57 (0.03) & \textbf{0.20} (0.04) & 0.13 (0.04)/0.07 (0.02)\\ 
$250, 25$ & \textbf{0.16} (0.07) & 0.33 (0.07)/0.30 (0.03) & 0.56 (0.04) & \textbf{0.26} (0.09) & 0.19 (0.05)/0.08 (0.03)\\ 
        \midrule
        \bottomrule
     \end{tabular}
  \end{threeparttable}}
\end{table}

\noindent The results from Table \ref{tab:diffk} indicate that the proposed model remains competitive for different $K$. One thing of note is that increasing the value of $K$ does not necessarily improve parameter estimates, even though there are more observations in total that can be used to improve parameter estimation. The reason is probably related to the fact that we only use a single fusion penalty parameter $\lambda_f$ rather than one penalty parameter per combination of groups $k$ and $k'$, with $k \neq k'$ whose coefficients are fused, such that our fusion approach reflects an averaged fusion, rather than a group specific one if $K > 2$.

\begin{table}[H]
\centering
\resizebox{0.8\textwidth}{!}{%
  \begin{threeparttable}
  \caption{Results for the fitted ROL models on simulated data with $\delta = 0.25, \eta = 0.8$ and $K = 4$. The RMSE (root mean square error), $F_1$ and RCR (rank correctness ratio) scores are averaged across 50 fitted models for each parameter combination and rounded to two decimals. The same holds for the RCR pred, which indicated the RCR on the aggregated list of the 20 alternatives with five new alternatives, whose rankings are predicted. Standard errors are provided between parentheses. Bold values represent the best result for that particular parameter combination across the four methods. SFROL represents the proposed method, whilst ROL and PROL represent the regular and pooled ROL methods respectively. Both metrics are provided for the SFROL, whilst only the RMSE is provided for the ROL and PROL methods, as these do not allow for variable selection}
  \label{tab:diffm}
     \begin{tabular}{c | cc | c | cc}
        \toprule
        \midrule
         & \multicolumn{5}{c|}{$M = 10, m = 5$}\\ \midrule
         & \multicolumn{2}{c|}{RMSE} & $F_1$ & \multicolumn{2}{c|}{RCR}\\ \midrule
         \textbf{$n_k, p$} & \textbf{SFROL} & \textbf{ROL/PROL} & \textbf{SFROL} & \textbf{SFROL} & \textbf{ROL/PROL}\\ \midrule
$25, 5$ & \textbf{0.17} (0.07) & 0.23 (0.08)/0.23 (0.07) & 0.54 (0.07) & \textbf{0.36} (0.18) & 0.28 (0.12)/0.22 (0.11)\\ 
$50, 5$ & \textbf{0.12} (0.05) & 0.15 (0.03)/0.21 (0.07) & 0.53 (0.09) & \textbf{0.48} (0.17) & 0.35 (0.10)/0.21 (0.11)\\ 
$100, 5$ & \textbf{0.11} (0.05) & 0.11 (0.04)/0.21 (0.07) & 0.54 (0.09) & \textbf{0.56} (0.18) & 0.46 (0.12)/0.25 (0.11)\\ 
$250, 5$ & \textbf{0.07} (0.04) & 0.07 (0.02)/0.20 (0.07) & 0.56 (0.10) & \textbf{0.64} (0.19) & 0.53 (0.15)/0.26 (0.13)\\ 
$25, 10$ & \textbf{0.29} (0.23) & 0.63 (0.40)/0.29 (0.10) & 0.53 (0.05) & \textbf{0.26} (0.11) & 0.20 (0.09)/0.16 (0.06)\\ 
$50, 10$ & \textbf{0.14} (0.07) & 0.34 (0.14)/0.25 (0.06) & 0.54 (0.09) & \textbf{0.39} (0.16) & 0.27 (0.09)/0.18 (0.10)\\ 
$100, 10$ & \textbf{0.14} (0.07) & 0.26 (0.10)/0.23 (0.06) & 0.55 (0.06) & \textbf{0.43} (0.17) & 0.31 (0.11)/0.18 (0.09)\\ 
$250, 10$ & \textbf{0.11} (0.06) & 0.18 (0.05)/0.22 (0.06) & 0.57 (0.06) & \textbf{0.53} (0.20) & 0.40 (0.12)/0.20 (0.08)\\ 
        \midrule
         & \multicolumn{5}{c|}{$M = 10, m = 5$}\\ \midrule
         & \multicolumn{2}{c|}{RMSE} & $F_1$ & \multicolumn{2}{c|}{RCR}\\ \midrule
         \textbf{$n_k, p$} & \textbf{SFROL} & \textbf{ROL/PROL} & \textbf{SFROL} & \textbf{SFROL} & \textbf{ROL/PROL}\\ \midrule
$25, 5$ & \textbf{0.10} (0.03) & 0.14 (0.04)/0.21 (0.07) & 0.51 (0.07) & \textbf{0.46} (0.13) & 0.37 (0.08)/0.21 (0.11)\\ 
$50, 5$ & \textbf{0.08} (0.03) & 0.09 (0.03)/0.21 (0.08) & 0.54 (0.07) & \textbf{0.60} (0.19) & 0.52 (0.17)/0.24 (0.13)\\ 
$100, 5$ & \textbf{0.06} (0.02) & 0.06 (0.02)/0.21 (0.07) & 0.56 (0.13) & \textbf{0.68} (0.18) & 0.57 (0.15)/0.24 (0.11)\\ 
$250, 5$ & 0.05 (0.04) & \textbf{0.04} (0.01)/0.20 (0.07) & 0.55 (0.10) & \textbf{0.72} (0.16) & 0.69 (0.10)/0.26 (0.13)\\ 
$25, 10$ & \textbf{0.19} (0.12) & 0.33 (0.18)/0.27 (0.11) & 0.52 (0.05) & \textbf{0.36} (0.18) & 0.29 (0.13)/0.16 (0.09)\\ 
$50, 10$ & \textbf{0.13} (0.13) & 0.24 (0.12)/0.23 (0.06) & 0.54 (0.05) & \textbf{0.50} (0.20) & 0.34 (0.11)/0.18 (0.08)\\ 
$100, 10$ & \textbf{0.09} (0.04) & 0.16 (0.06)/0.23 (0.06) & 0.54 (0.04) & \textbf{0.56} (0.18) & 0.41 (0.13)/0.19 (0.09)\\ 
$250, 10$ & \textbf{0.09} (0.06) & 0.13 (0.05)/0.22 (0.06) & 0.58 (0.06) & \textbf{0.63} (0.17) & 0.49 (0.13)/0.18 (0.10)\\ 
        \midrule
        \bottomrule
     \end{tabular}
  \end{threeparttable}}
\end{table}

\noindent When more alternatives are compared simultaneously in the partial rankings, the performance of all three approaches increase, as shown in Table \ref{tab:diffm}. Even though the number of observations does not increase, the observations themselves become more informative. Nevertheless, the proposed method still outperforms the alternative approaches in almost all cases for the RMSE and all cases for the RCR.

\begin{table}[H]
\centering
\scalebox{0.6}{%
  \begin{threeparttable}
  \caption{Results for the fitted ROL models on simulated data with $K = 4$, $m = 3$, $M = 20$ if $p < 25$ and $M = p$ otherwise. The RMSE (root mean square error) are averaged across 50 fitted models for each parameter combination and rounded to two decimals. Standard deviations are provided between parentheses. Bold values represent the best result for that particular parameter combination across the three methods. SFROL represents the proposed method, whilst ROL and PROL represent the regular and pooled ROL methods respectively.}
  \label{tab:simresbadrmse}
     \begin{tabular}{c | cc | cc | cc}
        \toprule
        \midrule
         & \multicolumn{2}{c|}{$\delta = 0, \eta = 0$} & \multicolumn{2}{c|}{$\delta = 0, \eta = 0.2$} & \multicolumn{2}{c}{$\delta = 0, \eta = 0.8$}\\ \midrule
         & \multicolumn{2}{c|}{RMSE} & \multicolumn{2}{c|}{RMSE} & \multicolumn{2}{c|}{RMSE} \\ \midrule
         \textbf{$n_k, p$} & \textbf{SFROL} & \textbf{ROL/PROL} & \textbf{SFROL} & \textbf{ROL/PROL} & \textbf{SFROL} & \textbf{ROL/PROL}\\ \midrule
$25, 5$ & 0.21 (0.10) & 0.36 (0.10)/\textbf{0.14} (0.05) & 0.17 (0.07) & 0.33 (0.09)/\textbf{0.14} (0.06) & 0.12 (0.06) & 0.28 (0.07)/\textbf{0.11} (0.04)\\ 
$50, 5$ & 0.19 (0.09) & 0.23 (0.09)/\textbf{0.09} (0.04) & 0.17 (0.08) & 0.23 (0.06)/\textbf{0.11} (0.04) & 0.10 (0.05) & 0.17 (0.04)/\textbf{0.07} (0.03)\\ 
$100, 5$ & 0.12 (0.06) & 0.14 (0.03)/\textbf{0.07} (0.03) & 0.10 (0.04) & 0.15 (0.04)/\textbf{0.07} (0.02) & 0.07 (0.04) & 0.12 (0.03)/\textbf{0.06} (0.02)\\ 
$250, 5$ & 0.09 (0.06) & 0.09 (0.02)/\textbf{0.04} (0.01) & 0.07 (0.03) & 0.09 (0.02)/\textbf{0.05} (0.02) & 0.06 (0.04) & 0.08 (0.02)/\textbf{0.04} (0.01)\\ 
$25, 10$ & \textbf{0.25} (0.09) & 0.99 (0.51)/0.26 (0.12) & 0.24 (0.09) & 0.82 (0.42)/\textbf{0.19} (0.08) & \textbf{0.12} (0.05) & 0.51 (0.21)/0.16 (0.06)\\ 
$50, 10$ & 0.21 (0.14) & 0.43 (0.18)/\textbf{0.14} (0.06) & 0.20 (0.07) & 0.36 (0.1)/\textbf{0.14} (0.05) & \textbf{0.09} (0.04) & 0.26 (0.06)/0.11 (0.03)\\ 
$100, 10$ & 0.14 (0.09) & 0.25 (0.07)/\textbf{0.10} (0.03) & 0.14 (0.05) & 0.22 (0.05)/\textbf{0.11} (0.03) & 0.08 (0.04) & 0.16 (0.03)/\textbf{0.07} (0.02)\\ 
$250, 10$ & 0.11 (0.06) & 0.13 (0.03)/\textbf{0.06} (0.02) & 0.10 (0.06) & 0.13 (0.03)/\textbf{0.07} (0.03) & 0.05 (0.03) & 0.10 (0.03)/\textbf{0.05} (0.02)\\ 
$25, 25$ & \textbf{0.43} (0.19) & 1.09 (0.23)/1.35 (1.09) & \textbf{0.36} (0.12) & 1.14 (0.28)/1.34 (0.81) & \textbf{0.18} (0.08) & 1.46 (0.34)/0.86 (0.64)\\ 
$50, 25$ & \textbf{0.34} (0.07) & 1.19 (0.25)/1.18 (0.83) & \textbf{0.31} (0.08) & 1.17 (0.33)/0.80 (0.52) & \textbf{0.15} (0.05) & 1.24 (0.36)/0.43 (0.28)\\ 
$100, 25$ & \textbf{0.31} (0.07) & 0.85 (0.21)/0.69 (0.42) & \textbf{0.26} (0.05) & 0.84 (0.24)/0.52 (0.44) & \textbf{0.12} (0.05) & 0.63 (0.15)/0.29 (0.13)\\ 
$250, 25$ & \textbf{0.25} (0.09) & 0.48 (0.12)/0.34 (0.19) & \textbf{0.23} (0.09) & 0.42 (0.13)/0.34 (0.21) & \textbf{0.09} (0.04) & 0.35 (0.10)/0.17 (0.06)\\ 
       \midrule
         & \multicolumn{2}{c|}{$\delta = 1, \eta = 0$} & \multicolumn{2}{c|}{$\delta = 1, \eta = 0.2$} & \multicolumn{2}{c}{$\delta = 1, \eta = 0.8$}\\ \midrule
         & \multicolumn{2}{c|}{RMSE} & \multicolumn{2}{c|}{RMSE} & \multicolumn{2}{c|}{RMSE} \\ \midrule
         \textbf{$n_k, p$} & \textbf{SFROL} & \textbf{ROL/PROL} & \textbf{SFROL} & \textbf{ROL/PROL} & \textbf{SFROL} & \textbf{ROL/PROL}\\ \midrule
$25, 5$ & \textbf{0.30} (0.09) & 0.35 (0.10)/0.48 (0.08) & \textbf{0.29} (0.06) & 0.31 (0.11)/0.49 (0.05) & \textbf{0.27} (0.07) & 0.38 (0.25)/0.46 (0.07)\\ 
$50, 5$ & \textbf{0.21} (0.06) & 0.22 (0.06)/0.47 (0.09) & \textbf{0.22} (0.07) & 0.24 (0.06)/0.49 (0.05) & \textbf{0.21} (0.06) & 0.22 (0.04)/0.45 (0.06)\\ 
$100, 5$ & 0.18 (0.04) & \textbf{0.15} (0.03)/0.47 (0.08) & 0.17 (0.04) & \textbf{0.16} (0.03)/0.48 (0.05) & 0.16 (0.04) & \textbf{0.15} (0.03)/0.44 (0.06)\\ 
$250, 5$ & 0.13 (0.07) & \textbf{0.09} (0.01)/0.47 (0.08) & 0.14 (0.05) & \textbf{0.08} (0.02)/0.48 (0.05) & 0.12 (0.05) & \textbf{0.09} (0.01)/0.44 (0.06)\\ 
$25, 10$ & \textbf{0.48} (0.39) & 0.92 (0.48)/0.55 (0.05) & \textbf{0.38} (0.07) & 0.92 (0.34)/0.52 (0.05) & \textbf{0.38} (0.19) & 0.91 (0.45)/0.49 (0.04)\\ 
$50, 10$ & \textbf{0.33} (0.08) & 0.40 (0.13)/0.54 (0.04) & \textbf{0.30} (0.08) & 0.39 (0.11)/0.51 (0.05) & \textbf{0.30} (0.07) & 0.35 (0.05)/0.47 (0.04)\\ 
$100, 10$ & 0.25 (0.05) & \textbf{0.24} (0.05)/0.53 (0.04) & \textbf{0.21} (0.06) & 0.22 (0.06)/0.50 (0.05) & 0.25 (0.07) & \textbf{0.21} (0.05)/0.47 (0.05)\\ 
$250, 10$ & 0.16 (0.04) & \textbf{0.13} (0.03)/0.53 (0.05) & 0.16 (0.05) & \textbf{0.13} (0.04)/0.50 (0.05) & 0.16 (0.04) & \textbf{0.12} (0.02)/0.47 (0.05)\\ 
$25, 25$ & \textbf{0.62} (0.23) & 1.07 (0.18)/0.97 (0.46) & \textbf{0.74} (0.31) & 1.11 (0.21)/1.03 (0.69) & \textbf{0.52} (0.16) & 1.10 (0.17)/0.91 (0.48)\\ 
$50, 25$ & \textbf{0.64} (0.29) & 1.00 (0.10)/0.77 (0.32) & \textbf{0.49} (0.28) & 1.13 (0.32)/0.67 (0.22) & \textbf{0.51} (0.25) & 0.98 (0.13)/0.54 (0.08)\\ 
$100, 25$ & \textbf{0.52} (0.25) & 0.86 (0.20)/0.62 (0.13) & \textbf{0.48} (0.23) & 0.78 (0.16)/0.66 (0.29) & \textbf{0.39} (0.15) & 0.78 (0.16)/0.52 (0.07)\\ 
$250, 25$ & \textbf{0.40} (0.08) & 0.46 (0.08)/0.59 (0.10) & \textbf{0.38} (0.08) & 0.43 (0.08)/0.56 (0.07) & \textbf{0.36} (0.09) & 0.43 (0.13)/0.52 (0.06)\\ 
        \midrule
        \bottomrule
     \end{tabular}
  \end{threeparttable}}
\end{table}

\noindent The result for the RMSE of the three approaches in the unfavourable simulations are similar to those of the RCR, see Tables \ref{tab:simresbadrmse} and \ref{tab:unfav} respectively, with the RCR results having a slight comparative edge for the proposed method. The reason for this is likely that the fitted models are optimised for rank prediction because of the cross-validation framework, which selects the combination of penalty parameters that maximise the RCR.
\\~\\
\noindent Instead of always assuming that $K$ is known in our simulations, we have also added some simulations containing a misspecified value of $K$. For these simulations, we either have that true $K = 2$ and assume that misspecified $\tilde{K} = 4$, or the other way around. If the number of assumed groups is incorrect, this not only influences the proposed method, but also influences the approach where we fit the ROL model separately on each group.

\begin{table}[H]
\centering
\resizebox{0.65\textwidth}{!}{%
  \begin{threeparttable}
  \caption{Results for the fitted ROL models on simulated data with $\delta = 0.25, \eta = 0.8$, $M = 20$ and $m = 3$. The RMSE (root mean square error), $F_1$ and RCR (rank correctness ratio) scores are averaged across 50 fitted models for each parameter combination and rounded to two decimals. The same holds for the RCR pred, which indicated the RCR on the aggregated list of the 20 alternatives with five new alternatives, whose rankings are predicted. Standard errors are provided between parentheses. Bold values represent the best result for that particular parameter combination across the three methods. SFROL represents the proposed method, whilst ROL and PROL represent the regular and pooled ROL methods respectively. Both metrics are provided for the SFROL, whilst only the RMSE is provided for the ROL and ROL methods, as these do not allow for variable selection}
  \label{tab:misspec}
     \begin{tabular}{c | cc | c | cc}
        \toprule
        \midrule
         & \multicolumn{5}{c|}{$K = 4, \tilde{K} = 2$}\\ \midrule
         & \multicolumn{2}{c|}{RMSE} & $F_1$ & \multicolumn{2}{c|}{RCR}\\ \midrule
         \textbf{$n_k, p$} & \textbf{SFROL} & \textbf{ROL/PROL} & \textbf{SFROL} & \textbf{SFROL} & \textbf{ROL/PROL}\\ \midrule
$25, 5$ & \textbf{0.23} (0.05) & 0.25 (0.05)/0.23 (0.07) & 0.53 (0.08) & \textbf{0.18} (0.14) & 0.12 (0.06)/0.16 (0.08)\\ 
$50, 5$ & \textbf{0.21} (0.06) & 0.21 (0.06)/0.22 (0.07) & 0.52 (0.09) & \textbf{0.23} (0.14) & 0.16 (0.07)/0.14 (0.08)\\ 
$100, 5$ & 0.20 (0.06) & \textbf{0.19} (0.06)/0.21 (0.07) & 0.56 (0.07) & \textbf{0.28} (0.20) & 0.18 (0.10)/0.16 (0.11)\\ 
$250, 5$ & 0.19 (0.07) & \textbf{0.18} (0.06)/0.20 (0.07) & 0.61 (0.13) & \textbf{0.26} (0.19) & 0.20 (0.11)/0.16 (0.11)\\ 
$25, 10$ & \textbf{0.24} (0.05) & 0.30 (0.07)/0.25 (0.05) & 0.53 (0.05) & \textbf{0.15} (0.07) & 0.13 (0.06)/0.11 (0.06)\\ 
$50, 10$ & \textbf{0.21} (0.04) & 0.23 (0.05)/0.22 (0.05) & 0.53 (0.06) & \textbf{0.17} (0.10) & 0.16 (0.08)/0.14 (0.08)\\ 
$100, 10$ & \textbf{0.18} (0.05) & 0.18 (0.04)/0.19 (0.04) & 0.54 (0.06) & \textbf{0.20} (0.08) & 0.19 (0.08)/0.16 (0.07)\\ 
$250, 10$ & 0.19 (0.05) & \textbf{0.17} (0.04)/0.20 (0.04) & 0.55 (0.06) & \textbf{0.23} (0.16) & 0.19 (0.09)/0.16 (0.07)\\ 
$25, 25$ & \textbf{0.39} (0.29) & 1.89 (1.12)/0.76 (0.47) & 0.52 (0.03) & \textbf{0.08} (0.02) & 0.06 (0.03)/0.07 (0.02)\\ 
$50, 25$ & \textbf{0.27} (0.04) & 0.87 (0.41)/0.44 (0.17) & 0.53 (0.05) & \textbf{0.10} (0.04) & 0.07 (0.02)/0.07 (0.03)\\ 
$100, 25$ & \textbf{0.31} (0.15) & 0.51 (0.32)/0.47 (0.24) & 0.54 (0.04) & \textbf{0.09} (0.04) & 0.06 (0.02)/0.08 (0.04)\\ 
$250, 25$ & \textbf{0.24} (0.04) & 0.35 (0.12)/0.38 (0.17) & 0.56 (0.04) & \textbf{0.11} (0.04) & 0.09 (0.03)/0.10 (0.03)\\ 
        \midrule
         & \multicolumn{5}{c|}{$K = 2, \tilde{K} = 4$}\\ \midrule
         & \multicolumn{2}{c|}{RMSE} & $F_1$ & \multicolumn{2}{c|}{RCR}\\ \midrule
         \textbf{$n_k, p$} & \textbf{SFROL} & \textbf{ROL/PROL} & \textbf{SFROL} & \textbf{SFROL} & \textbf{ROL/PROL}\\ \midrule
$25, 5$ & 0.25 (0.10) & 0.43 (0.11)/\textbf{0.21} (0.08) & 0.48 (0.11) & \textbf{0.18} (0.20) & 0.10 (0.07)/0.16 (0.09)\\ 
$50, 5$ & 0.23 (0.09) & 0.35 (0.11)/\textbf{0.18} (0.08) & 0.48 (0.11) & \textbf{0.20} (0.24) & 0.12 (0.05)/0.18 (0.08)\\ 
$100, 5$ & 0.22 (0.09) & 0.28 (0.10)/\textbf{0.16} (0.07) & 0.43 (0.07) & \textbf{0.22} (0.14) & 0.14 (0.07)/0.17 (0.10)\\ 
$250, 5$ & 0.21 (0.10) & 0.23 (0.10)/\textbf{0.14} (0.08) & 0.50 (0.14) & \textbf{0.29} (0.17) & 0.19 (0.09)/0.22 (0.14)\\ 
$25, 10$ & 0.27 (0.07) & 0.56 (0.17)/\textbf{0.25} (0.06) & 0.50 (0.07) & 0.13 (0.09) & 0.08 (0.05)/\textbf{0.14} (0.07)\\ 
$50, 10$ & 0.24 (0.07) & 0.50 (0.15)/\textbf{0.22} (0.06) & 0.51 (0.07) & 0.15 (0.09) & 0.11 (0.04)/\textbf{0.16} (0.07)\\ 
$100, 10$ & 0.21 (0.08) & 0.31 (0.08)/\textbf{0.16} (0.05) & 0.49 (0.07) & \textbf{0.22} (0.12) & 0.13 (0.07)/0.21 (0.12)\\ 
$250, 10$ & 0.17 (0.07) & 0.22 (0.05)/\textbf{0.13} (0.05) & 0.48 (0.04) & \textbf{0.27} (0.16) & 0.17 (0.07)/0.26 (0.14)\\ 
$25, 25$ & \textbf{0.34} (0.12) & 1.68 (0.42)/0.78 (0.71) & 0.48 (0.04) & \textbf{0.08} (0.05) & 0.05 (0.03)/0.07 (0.04)\\ 
$50, 25$ & \textbf{0.31} (0.12) & 1.57 (0.36)/0.74 (0.69) & 0.49 (0.04) & \textbf{0.11} (0.05) & 0.08 (0.02)/0.09 (0.04)\\ 
$100, 25$ & \textbf{0.34} (0.16) & 1.37 (0.36)/0.45 (0.11) & 0.47 (0.05) & \textbf{0.10} (0.03) & 0.07 (0.02)/0.10 (0.06)\\ 
$250, 25$ & \textbf{0.22} (0.06) & 0.65 (0.18)/0.27 (0.06) & 0.48 (0.04) & \textbf{0.17} (0.08) & 0.09 (0.03)/0.12 (0.07)\\ 
        \midrule
        \bottomrule
     \end{tabular}
  \end{threeparttable}}
\end{table}

\noindent Unsurprisingly, model performance for both the proposed method and the separate ROL method suffer substantially when $K$ is misspecified, as judged from the results in Figure \ref{fig:boxplots} and Table \ref{tab:simres1} versus those of Table \ref{tab:misspec}. Nevertheless, the proposed method still manages to outperform the alternative approaches in most cases, both in terms of the RMSE and the RCR. This is an interesting result, as the pooled ROL approach is expected to perform best, given that it is the only approach we evaluate that is not affected by a misspecified $K$. It appears that the gain from imposing sparsity and fusion is bigger than the loss incurred by the misspecification of $K$. Similar to the other simulated data, performance is dependent on both the sparsity and heterogeneity of the coefficients, as well as the dimensionality of the data. Whilst the pooled approach has a natural advantage for misspecified $K$ over the proposed method, for large enough $p$ and with sufficiently sparse coefficients that exhibit some heterogeneity, the proposed method is expected to outperform the alternative approaches.

\section*{Appendix B: Identifiability}
\noindent A model is identifiable if and only if for parameters $\theta_1, \theta_2 \in \Theta$, with $\theta_1 \neq \theta_2$ and for a given set of samples $\bm{X} \in \mathcal{X}$, we have that $L(\theta_1) \neq L(\theta_2)$, where $L(\cdot)$ denotes a likelihood function. Here we show that in order for the ROL model to be identifiable, we require that rank$(\bm{X}) = p$. 
\begin{proof}
\noindent Let $\bm{\Pi}^{(1)},\ldots,\bm{\Pi}^{(K)}$,  $\bm{\Pi}^{(k)} \in \mathbb{N}^{n_{k} \times m}$ and $\bm{X} \in \mathbb{R}^{M \times p}$ be given. If, for some $\bm{B}_1 = \left\{\bm{\beta}_1^{(1)}, \ldots, \bm{\beta}_1^{(K)}\right\}, \bm{B}_2\ = \left\{\bm{\beta}_2^{(1)}, \ldots, \bm{\beta}_2^{(K)}\right\}$, with $\bm{\beta}_1^{(k)}, \bm{\beta}_2^{(k)} \in \mathbb{R}^{p}$ we have that $L(\bm{B}_1) = L(\bm{B}_2)$, then
\begin{equation*}
\begin{gathered}
    \prod_{k = 1}^K \prod_{i = 1}^{n_k} \prod_{j = 1}^{m}\frac{\exp\left(\bm{x}_{\sigma^{(k)}_{ij}}\bm{\beta}_1^{(k)}\right)}{\sum_{l = j}^m \exp\left(\bm{x}_{\sigma^{(k)}_{il}}\bm{\beta}_1^{(k)}\right)} = \prod_{k = 1}^K \prod_{i = 1}^{n_k} \prod_{j = 1}^{m}\frac{\exp\left(\bm{x}_{\sigma^{(k)}_{ij}}\bm{\beta}_2^{(k)}\right)}{\sum_{l = j}^m \exp\left(\bm{x}_{\sigma^{(k)}_{il}}\bm{\beta}_2^{(k)}\right)}\\
     \sum_{k=1}^K\sum_{i = 1}^{n_k}\sum_{j = 1}^{m}\log\left[\frac{\sum_{l = j}^m \exp\left(\bm{x}_{\sigma^{(k)}_{il}}\bm{\beta}_2^{(k)}\right)}{\sum_{l = j}^m \exp\left(\bm{x}_{\sigma^{(k)}_{il}}\bm{\beta}_1^{(k)}\right)}\right] + \sum_{k=1}^K\sum_{i = 1}^{n_k}\sum_{j = 1}^{m}\bm{x}_{\sigma^{(k)}_{ij}}\left(\bm{\beta}_1^{(k)} - \bm{\beta}_2^{(k)}\right) = 0,
    \end{gathered}
\end{equation*}
iff $\bm{X}\bm{\beta}_1^{(k)} = \bm{X}\bm{\beta}_2^{(k)}$ for all $k$. This is true irrespective of the rank of $\bm{X}$, whenever $\bm{\beta}_1^{(k)} = \bm{\beta}_2^{(k)}$ for all $k$. However, note that if rank$(\bm{X}) < p$, then it is well known that for any linear system of equations, there exists $\bm{\beta}_1 \neq \bm{\beta}_2$ whilst $\bm{X}\bm{\beta}_1^{(k)} = \bm{X}\bm{\beta}_2^{(k)}$. Therefore, we require that rank$(\bm{X}) = p \leq M$.
\end{proof}

\bibliographystyle{Chicago}
\bibliography{library}

\end{document}